\documentclass{elsarticle}
\usepackage{epsfig}
\usepackage{a4wide}

\def\lb#1{\if 1#1 \ln\beta \else \ln^#1\beta \fi}
\def\lt#1{\if 1#1 \ln 2 \else \ln^#1 2 \fi}

\newcommand{\be}{\begin{equation}}
\newcommand{\ee}{\end{equation}}
\newcommand{\ba}{\begin{eqnarray}}
\newcommand{\ea}{\end{eqnarray}}

\def\vec#1{{\mbox{\boldmath$#1$}}}

\newcommand{\gsim}{\mbox{\raisebox{-0.3ex}{%
\footnotesize $\:\stackrel{>}{\sim}\:$}} }

\def\B{{\cal B}}
\def\tree{{\rm tree}}
\newcommand{\qb}{{\bar q}}

\newcommand{\tb}{{\bar t}}
\newcommand{\nn}{\nonumber}
\newcommand{\Fslash}[1]{#1 \hspace{-0.42em}/\hspace{-0.08em}}

\def\ib{{\bar\imath}}
\def\oneloop{{1 \mbox{-} \rm loop}}

\begin{document}


 \vspace{\baselineskip}

\title{
NLO QCD corrections to top quark pair production in association
with one hard jet at hadron colliders
}

    \author{Kirill Melnikov and Markus Schulze}
    \address{
Department of Physics and Astronomy,
Johns Hopkins University,
Baltimore, MD, USA}


    \begin{abstract}
      \noindent
We compute the QCD corrections to the production of
a $t \bar t$  pair in association with one hard jet
at the Tevatron and the LHC, using the method of generalized
$D$-dimensional unitarity.  Top quark decays
are included  at leading order in perturbative QCD.
We present kinematic distributions of top quark decay products
in lepton plus jets  and dilepton final states at the Tevatron
and the LHC,
using realistic selection cuts.
We confirm a strong  reduction of the top quark forward-backward
asymmetry
in $p \bar p \to t \bar t j$ at next-to-leading order, first
observed by Dittmaier, Uwer and Weinzierl.  We argue that there is
a natural way to understand this reduction and that it does not
imply a breakdown of the perturbative expansion for the asymmetry.
    \end{abstract}

    \maketitle

\section{Introduction}

Experiments at the Large Hadron Collider (LHC) are in the process
of testing  the very successful paradigm
of the past thirty years -- the Standard Model of particle physics
-- in an entirely   new energy regime. The goal of these experiments
is to unravel  the mechanism of  electroweak symmetry breaking.
Electroweak symmetry breaking may
occur due to the Higgs mechanism, as in the Standard Model,
or it may be a consequence of more complex dynamics.
In the Standard Model the unusually  large mass of the top quark  is
introduced by hand and top quarks do not play any role  in electroweak
symmetry breaking. This may change in extensions of the Standard Model.
Moreover,  the large mass of the top quark makes its
interactions with an agent of electroweak symmetry breaking parametrically
enhanced, relative to other quarks. Therefore,
top quarks provide a window of opportunity to investigate physics
of electroweak symmetry breaking; this motivates a rigorous
experimental program to study  top quarks and their  properties.

More than  a decade after the discovery of the top quark in
1995 \cite{disc_cdf,disc_d0},
the Tevatron remains  at the forefront of  top quark studies.
During that time,  CDF and D0 reached
remarkable  precision in measuring the top quark production cross-section in
$p \bar p$ collisions and  the top quark  mass \cite{tmass}.
 Also,  the top quark
charge  is constrained by
experimental measurements \cite{tcharge_cdf,tcharge_d0}
and the hypothesis that the top quark spin is one half follows
indirectly from the $p \bar p \to t \bar t$ cross-section measurement
and is consistent with the observation of the top quark decay $t \to Wb$.
The structure and the
strength of $t bW$ interaction vertex was thoroughly studied
\cite{tbw_cdf,tbw_d0}
and limits on  exotic  contributions to
top quark decays were placed.
A recent  measurement  of the forward-backward
asymmetry  of the top quark \cite{ass_cdf,ass_d0}
may point towards non-standard
contributions to top quark
pair production;
currently, such a possibility is being   actively
discussed.

In spite of the enormous integrated luminosity accumulated  by the
Tevatron, many of their measurements are limited by statistics since
the ratio of the hadron center of mass collision energy
to twice the top quark mass is not large.
This will change with the start-up of the LHC since the LHC
will be  a $t \bar t$ factory. It is
expected that many of the  studies performed at the Tevatron
will be
continued and extended at the LHC
with much higher statistics.
The higher energy of the LHC leads to a much
larger cross-section  for the
top quark pair production, compared to the Tevatron,
and, also,  to a greater  variety of
 the kinematics of the produced top quarks.
In contrast to
the Tevatron, top quarks are produced with larger energies and transverse
momenta at the LHC. This kinematic feature implies  larger
probability for top quarks to radiate gluons
and the increase of the relative
importance of the $t \bar t + {\rm jets}$ final state in the inclusive
$t \bar t$ sample.

It is important to understand how many jets are produced in association
with a $t \bar t$ pair.  Indeed, since
top quarks may decay into leptons, missing energy and jets,
$t \bar t$  production is an important  background
to standard signatures of physics beyond the Standard Model, where
the number of jets in the process is often taken as a
discriminator against the background.
Note also that selection cuts that enhance New Physics
contribution to a particular observable typically select harder top quarks,
thereby enhancing the probability of QCD radiation.
An interesting example is the study of $t \bar t$
background to the Higgs boson production in vector boson fusion
$pp \to H + 2~{\rm jets} \to W^+W^- + 2~{\rm jets}
\to l \bar l \nu \bar \nu + 2~{\rm jets}$. The final
state is identical to the $ t \bar t$ production if
both top quarks decay leptonically and the two jets are associated with
$b$-jets from top quark decays. Requirement that the
two jets are
separated by a large rapidity interval
suppresses  the $pp \to t \bar t$ background since $b$-jets from
top quark decays
tend to be central. On the other hand,
in $ pp \to t \bar t +{\rm jet}$ process there exists
kinematic configuration with
one central, one forward and one backward
jet, separated by large rapidity gap. Since one of
the forward/backward  jets can be a gluon jet from the initial state radiation,
only one $b$-jet from top decays has to be non-central. Because
the probability to have one $b$-jet from top
decays going in  the forward/backward
direction is much higher than the probability that $b$-jets from
{\it both}
top decays are in the forward/backward direction,
$t \bar t + {\rm jet}$ becomes  the dominant background.
There are other cases in which
$t \bar t + {\rm jets}$ background is important and $t \bar t + {\rm jet}$
is an essential part of it. For example, the importance
of the $t \bar t + {\rm jets}$ background {\it increases} with the
number of jets \cite{mangano},
 for  supersymmetry
 searches in events with jets + missing  energy.

Accurate theoretical description of $pp(p \bar p)
\to t \bar t + {\rm jets}$ process
requires next-to-leading order (NLO) QCD calculations. Such calculations
are difficult because of the large number of diagrams and the
presence of  massive particles. While NLO QCD corrections to
$t \bar t$ production are known since early 1990s
\cite{nde,bnmss,mnr,fnmr},
a calculation of NLO QCD corrections to $t \bar t + {\rm jet}$
was reported  recently \cite{Dittmaier:2007wz, Dittmaier:2008uj}.
Even more  recently first results for NLO QCD corrections
to $pp \to t \bar t + 2~{\rm jets}$
were presented in Ref.~\cite{Bevilacqua:2010ve}. NLO QCD
corrections are also known for other associated production
processes that involve top quarks, including
$t \bar t \gamma$ \cite{ttg}, $t \bar t Z$ \cite{ttz}, $t \bar t H$ \cite{tth}
and $t \bar t b \bar b$ \cite{Bredenstein:2009aj,Bredenstein:2010rs,Bevilacqua:2009zn}.
However, quite often,
when NLO QCD corrections  are computed,
top quarks  are treated as stable
particles and their decays, even at leading order, are not included.
This is not entirely realistic since
all  cuts designed to either select a $t \bar t$ sample or to
reject it, to search for physics beyond the Standard Model,
apply  to  top quark decay products.  It is well-known that kinematics
of the decay products is affected by spin correlations. Therefore, it
may be   important to account for decays of top quarks in the
calculations of
NLO QCD  corrections, to have full confidence in the results. We note in
this regard that all acceptances employed in experimental studies of top
quarks convert a measurement
with cuts on the top quark decay products to a result that refers
to top quarks as stable particles, are currently calculated
with parton shower event generators
and are {\it never} corrected for higher order QCD effects.
This is equivalent to declaring, without any proof or argument,
that QCD radiative corrections to top quark pair production
are always reduced to kinematic-independent $K$-factors and, therefore,
cancel out in the calculation of all acceptances. Whether or not such
an assertion is correct, can only be established
through a next-to-leading order calculation that includes
top quark decay products.
We note in this
regard that  computation of NLO QCD corrections to observables
in $t \bar t$ pair production that are
sensitive to top quark decay products was pioneered
in Refs. \cite{Bernreuther:2001bx, Bernreuther:2001rq, Bernreuther:2001jb,
brand, Bernreuther:2004jv,Bernreuther:2004wz}.
The complete calculation of NLO QCD corrections to $t \bar t$ pair
production that retains spin correlations,  allows application of arbitrary
cuts to the decay products of top quarks and includes
QCD corrections to the  decay, was presented
very recently in Refs.~\cite{ms,bernreuther}.

In this paper, we extend the results  of
Refs.~\cite{Dittmaier:2007wz, Dittmaier:2008uj} by
computing NLO QCD corrections to  $ pp(p \bar p) \to t \bar t + {\rm jet}$,
including leptonic and hadronic decays of top quarks at leading order.
To calculate one-loop virtual amplitudes,
we employ the method  of generalized $D$-dimensional unitarity
suggested in Ref.~\cite{Giele:2008ve} and applied to massive
particles in Ref.~\cite{egkm}.  We point out that  a basic object that
needs to be calculated for each phase-space point
within the unitarity framework is a one-loop
{\it helicity amplitude}. Since helicity amplitudes need to
be calculated {\it anyhow},  it takes (roughly) the same effort to calculate
helicity   density matrix
$\rho_{\lambda \lambda'} \sim A_\lambda A^*_{\lambda'}$, needed for
an exact account of
top quark decays, and the spin-averaged cross-section
$\sigma \sim \sum_\lambda \rho_{\lambda \lambda}$. Therefore, when
unitarity methods are employed in the calculation of one-loop
corrections, there is no good reason to avoid including the leading order
on-shell top quark decay.  Note, however,
that in order to produce a result for $pp(p \bar p) \to t \bar t
+{\rm jet}$ where NLO QCD corrections to {\it both} production and decay are
included, one needs to know NLO QCD corrections to
the decay $t \to b W+{\rm jet}$.
Such calculation is  straightforward but tedious;
we hope to return to it
in a separate  publication.

The remainder of the paper is organized as follows. In Section~\ref{sect2}
we  comment on some aspects of generalized unitarity that
are important for the  calculation that we report in this paper.
The  goal is to provide details which were not presented clearly  in
the original
publications \cite{Giele:2008ve,egkm}.
In Section~\ref{sect3} technical aspects of the calculation
are reviewed.  In Section~\ref{sectr}
phenomenological results are described.
We conclude in Section~\ref{sectc}.

\section{Aspects of  generalized $D$-dimensional unitarity}
\label{sect2}

To calculate one-loop virtual corrections, we employ the
framework of generalized $D$-dimensional unitarity, following
Refs.~\cite{Giele:2008ve,egkm}. This method grew out of
the observation of Ref.~\cite{egk} that the   procedure
for tensor reduction of one-loop integrals, suggested in Ref.~\cite{opp},
meshes well with unitarity-based methods
 and allows computation of one-loop on-shell
scattering amplitudes, rather than Feynman diagrams.  The central
idea of Refs.~\cite{Giele:2008ve,egkm} is that this technique can be
extended in such a way that both, cut-constructible and rational
parts of the amplitude, can be obtained in an efficient way.

While Refs.~\cite{Giele:2008ve,egkm} explain clearly how
numerical computation can be set up, it is beneficial
to discuss some aspects of the implementation in more detail.
Recall that within the
context of generalized $D$-dimensional unitarity, we deal with
the  so-called primitive amplitudes \cite{bdk}, which are  gauge-invariant
subsets of color-ordered amplitudes. The importance of primitive
amplitudes for generalized unitarity follows from the fact that
external particles in a primitive amplitude are ordered and no permutations
are allowed. This feature enables enumeration of  all the propagators,
that contribute to such an amplitude, in an unique fashion\footnote{Recently,
a procedure that avoids the necessity to deal with  primitive
amplitudes in the context of unitarity-based one-loop
computations was described  in Ref.~\cite{gkw}.}.

As explained in Refs.~\cite{Giele:2008ve,egkm}, a full one-loop amplitude
can be computed if the {\it integrand} of the one-loop amplitude
is known in the extension of the four-dimensional
theory to $D > 4$ integer dimensions, and the loop momentum is restricted
to five dimensions. In this case,  an
integrand of any $N$-point primitive amplitude can be decomposed into
a sum of terms  with at most five Feynman propagators
\be
{\cal A}_N(l) =
\sum \limits_{[i_1|i_5]}^{} \frac{{\bar e}^{}_{i_1 i_2 i_3 i_4 i_5}}{d_{i_1} d_{i_2} d_{i_3} d_{i_4} d_{i_5}}
+
\sum \limits_{[i_1|i_4]}^{} \frac{{\bar d}^{}_{i_1 i_2 i_3 i_4}}{d_{i_1} d_{i_2} d_{i_3} d_{i_4}}
+
\sum \limits_{[i_1|i_3]}^{} \frac{{\bar c}^{}_{i_1 i_2 i_3}}{d_{i_1} d_{i_2} d_{i_3} }
+
\sum \limits_{[i_1|i_2]}^{} \frac{{\bar b}^{}_{i_1 i_2}}{d_{i_1} d_{i_2}}
+\sum \limits_{[i_1]}^{} \frac{{\bar a}^{}_{i_1}}{d_{i_1}}.
\label{eq21}
\ee
In Eq.(\ref{eq21}) we use the notation $[i_1|i_n] = 1 \le i_1<i_2<...<i_n\le N$ and
$d_i(l) = (l+p_i)^2 - m_i^2$, $i =1...N$.
The left hand side of Eq.(\ref{eq21}) is completely specified by
Feynman rules. The challenge is to find an efficient and numerically
stable way to find  the coefficients $\bar e^{}_{i_1..i_5},..,
\bar a^{}_{i_1}$
on  the right hand side of Eq.(\ref{eq21}).
It was suggested in Ref.~\cite{opp} that a very efficient way to do that
is to solve for the coefficients $\bar e^{}_{i_1..i_5},...,\bar a^{}_{i_1}$
sequentially, by calculating left and right hand sides of Eq.(\ref{eq21})
for special values of the loop momentum $l$, where  certain subsets of
inverse Feynman propagators $d_{1},d_{2},...,d_{N}$ vanish.
All propagator sets with up to five members should be considered.
It was shown  in Ref.~\cite{egk}, that the coefficients
$\bar e^{}_{i_1..i_5},...,\bar a^{}_{i_1}$ have  restricted functional
dependence on the loop momentum $l$;  they are
typically given  by linear combinations
of constant terms and traceless tensors of ranks up to
five, four, three, two and one, respectively.
Those tensors are defined on linear spaces constructed
from basis  vectors, orthogonal to momenta present in
Feynman propagators  of a corresponding term in Eq.(\ref{eq21}).
The dimensionality of these transverse spaces is
$D-4$, $D-3$, $D-2$, $D-1$, $D$
for terms with five, four, three,  two and one denominators,
respectively.  It can be shown that, after integration over the loop
momentum, the contribution of the traceless tensors vanishes.

To illustrate this, we consider a term with two denominators
$d_{i_1,i_2} = d_{1,2}$
in Eq.(\ref{eq21}).
We choose the momentum $l$ in such a way that $d_{1,2}$ read
\be
d_1 = l^2 - m_1^2,\;\;\; d_2 = (l+p)^2 - m_2^2.
\label{eq22}
\ee
We write $l$ as a linear combination of the vector $p$,
that enters $d_2$ and the vector $l_\perp$ that belongs
to the transverse space described earlier
\be
l^\mu = x p^\mu + l_\perp^\mu;\;\;\;\; p l_\perp =  0.
\label{eq222}
\ee
The numerator function $b_{12}(l)$ is given by
the sum of two terms -- an $l$-independent
constant $b_{12}^{(0)}$ and a function
$\tilde b_{12}(l_\perp)$
that depends on the transverse component of the momentum $l$
\be
b_{12}(l) = b_{12}^{(0)} + \tilde b_{12}(l_\perp).
\label{eqm124}
\ee
The integral of the function $b_{12}(l_\perp)$
over directions of the transverse space vanishes
\be
\int {\rm d}^{D-1}l_\perp \delta (l_\perp^2 - \mu_0^2)
\tilde b_{12}(l_\perp) = 0.
\label{eqm123}
\ee
Because $d_{1,2}$ depend on $l_\perp^2$, Eq.(\ref{eqm123}) implies
that $\tilde b_{12}(l_\perp)$ does not contribute to the final
result where one-loop amplitude is expressed through scalar
(master) integrals.  As explained  in Refs.~\cite{opp,egk,Giele:2008ve,egkm},
renormalizability of the theory and the fact that momenta of all
external
particles are kept in four-dimensions restricts the
functions
${\bar e}_{i_1...i_5},...,{\bar a}_{i_1}$.
For example, as we already mentioned, ranks of tensors that contribute
to those functions are limited  and, also,
those functions  must be parity even with respect
to  the fifth  component of the loop momentum $l_5$.

The general  set up that we just presented  is employed in many
existing applications of  generalized unitarity
\cite{ms,egkmz,gz,laz,laz1}. It
is discussed in  detail in the original papers~\cite{Giele:2008ve,egkm}. However, there are some aspects of the
implementation of this method that are not adequately described in
those references and that warrant some discussion.
This is what we do in the remainder of this Section.

Our first comment concerns the coefficient of the term with five denominators
$e_{i_1 i_2..i_5}$. According to the preceding discussion, it can be
any even tensor up to rank five, composed of $l_5$. Hence,
the most general form of this function is
\be
\bar e = e_0 + e_1 l_5^2 + e_2 l_5^4.
\ee
However, it is easy to see that
\be
l_5^2 = f_0 + {\cal O}(d_1,d_2,...d_5),\;\;\;
l_5^4 = f_0^2 + {\cal O}(d_1,d_2,...d_5),
\label{eq2}
\ee
where $f_0$ is an $l$-independent term that only depends
on the kinematic variables of a particular five-point function.
It follows from Eq.(\ref{eq2}) that either scalar five-point integral
or  the five-point integrals with  $l_5^2$ or $l_5^4$ in the
numerator
can be chosen to be the  five-point master integral since the
difference between these integrals is
given by linear combinations  of  the four-point functions.
In Ref.~\cite{Giele:2008ve}
the scalar five-point integral  was chosen as the master integral
but, as  was realized later \cite{egkmz}, that choice
is  somewhat unfortunate. Indeed,
it is well-known that  if one calculates the cut-constructible part
of a one-loop amplitude,
one does not need to consider the five-point master integrals. On the
other hand, if one chooses ${\bar e}$ to be
an $l$-independent  constant,  the scalar five-point
function appears as a master integral. Because in the $D \to 4$ limit
the five-point function is not independent of
the  four-point functions,  and because  this is
the limit of interest,  large numerical cancellations   between the
four-point  and the five-point master integrals  are to be expected
and indeed occur.
These cancellations may be so strong
that the accuracy of the result deteriorates.
To ensure that all numerical
cancellations happen {\it locally}, it is useful
to choose the basis of master integrals
such that  $e_{i_1..i_5} \sim l_5^2 \sim l_{D-4}^2$.
Since
\be
\lim_{D \to 4} \int \frac{{\rm d}^Dl}{(2\pi)^D}
\frac{l_{D-4}^2}{d_1 d_2 d_3 d_4 d_5} \to 0,
\ee
the new master integral does not contribute to the final result,
but it is needed at the intermediate stages for proper identification
of lower-point functions through the residues of the one-loop scattering
amplitude. The new choice
of the  master integral for the five-point function helps to avoid
large numerical cancellations. A similar discussion of the
role of the five-point master integral in the context of $D$-dimensional
unitarity was recently given in Ref.~\cite{laz}.

Next, we describe a procedure
to  find functions ${\bar e}_{i_1..i_5},...,{\bar a}_{i_1}$
in a numerically stable way. We put special emphasis on  exceptional
cases where the general reduction algorithm needs to be extended.
To simplify the discussion, we consider four dimensional unitarity and
focus on
terms with   two denominators in Eq.(\ref{eq21}).
We stress
that all the subtleties associated with the reduction
can be  illustrated  by those considerations.
To this end, we pick any  term with two denominators in Eq.(\ref{eq21});
we refer to the inverse Feynman propagators
in that term as $d_1$ and $d_2$ and
to its numerator as $b(l)$. For ease of the presentation,
we only discuss the cut-constructible part in what follows.
We assume that $d_1$ and $d_2$ and the momentum $l$
are given by Eqs.~(\ref{eq22},\ref{eq222})
and write $l_\perp$ as
\be
 l_\perp = \sum_{i=2}^{4} x_{i} v_i,\;\;\;
p v_i = 0,\;\;\;v_i v_j = \delta_{ij}.
\label{eq20}
\ee
In Eq.(\ref{eq20}), we introduced  basis vectors
$v_{i}$, $i=2,3,4$ for the
linear space which is transverse to the momentum $p$.
As explained in \cite{egk,egkm},
the function $b(l)$ can be  written as
\ba
&& {\bar  b}(l) = b_1+ b_2 (v_2 l) + b_3 (v_3 l)  + b_4 (v_4 l)
+ b_5 \left ( ( v_2 l)^2 - (v_3 l)^2 \right )
\nonumber \\
&&\;\;\;\;\;
+ b_6 \left ( ( v_2 l)^2 + (v_3 l)^2 - 2(v_4 l)^2 \right )
+ b_7 (v_2 l) ( v_3 l) + b_8 ( v_2 l) ( v_4 l) + b_9 ( v_3 l) ( v_4 l).
\label{eq145}
\ea
In Eq.(\ref{eq145}) $b_{1,.,9}$ are unknown $l$-independent
parameters; the goal is to set up an algorithm to find them.
To calculate $b_{1,.,9}$, we would like
to calculate $b(l)$ for some values of the
loop momentum and then use the results of the computation
to solve the system of linear
equations where $b_{1,.,9}$ are treated as the unknown parameters.

 It was suggested
in Ref.~\cite{opp}
that it is convenient to choose the  momentum $l$
for which both inverse
Feynman propagators vanish
$d_1(l) = d_2(l) = 0$. In Ref.~\cite{egk} it was
pointed out that, for such choices of the loop momenta, the
function $b(l)$ and all other numerator functions in Eq.(\ref{eq21})
can be calculated using tree-level on-shell scattering amplitudes.
This feature is apparent  from Eq.(\ref{eq21}) since
functions $\bar e, \bar d, \bar c, \bar b, \bar a$ are
the residues of one-loop scattering amplitudes in a situation where
certain virtual particles go on the mass-shell. Such residues are
computed by cutting some internal
lines in an one-loop amplitude; this
terminates the flow of the loop momentum and turns a one-loop amplitude
into a product of tree-level on-shell amplitudes.
These on-shell amplitudes are conventional apart from the fact
that they must be calculated for the external {\it complex}  on-shell
momentum. Using Eq.(\ref{eq20}), we find  that  components of the momentum
$l$ for which  $d_{1,2} = 0$ are subject to the following constraints
\be
x = \frac{( m_2^2 - m_1^2 - p^2)}{2p^2},\;\;\; l_\perp^2 = m_1^2 - x^2 p^2.
\label{eq23}
\ee
It follows from Eq.(\ref{eq23})  that there are infinitely many
momenta $l$   that satisfy the $d_{1,2}=0$  condition since
only $l_\perp^2$ is constrained. Therefore, all we need to do
is to choose nine vectors with
different directions of $l_\perp$, calculate  $b(l)$ for those
vectors and solve the resulting linear system of equations to find
$b_{1,.,9}$.

In principle, the nine vectors  can be chosen arbitrarily.  However,
a more systematic procedure emerges if we parameterize
the transverse component of the loop momentum as
\be
l_\perp = |l_\perp|
\left ( \sin \theta \cos \phi \; v_2
+ \sin \theta \sin \phi \; v_3
+  \cos \theta \; v_4 \right ).
\ee
When this parameterization is employed in Eq.(\ref{eq145}),
$b(l)$   becomes a degree-four
polynomial of  $e^{i \theta}, e^{i \phi}$. Since we are interested in
finding coefficients of that polynomial,  the technique of
discrete Fourier transform can be employed. In the context of unitarity
methods,  this technique was described
in Refs.~\cite{oppm,bh1} and we do not repeat it here.

Note, however, that the application of the discrete Fourier transform
requires  division by $|l_\perp|$.
According to Eq.(\ref{eq23}), $|l_\perp|$
vanishes if  $m_1^2 = x^2 p^2$
which corresponds to $ p^2 = (m_2-m_1)^2$ or $p^2 = (m_2 + m_1)^2$.
These kinematic points are not dangerous if only
massless virtual  particles are considered since the
offending integrals
are scale-less two-point functions that are  discarded in dimensional
regularization.   The situation changes once  virtual massive
particles are considered and this is the case that we are interested in.
Note also that close to those exceptional
values of $p^2$,  $|l_\perp|$ can be small, so that
division by it may lead to numerical instabilities.

To handle the  case of small $|l_\perp|$ in a numerically stable
way, the method of discrete Fourier transform  is not directly
applicable
and the system of equations must be  solved differently.
There are many ways to solve a system of linear equations
avoiding division   by $|l_\perp|$; the procedure
that we have implemented in the numerical program
is described below.
We begin by choosing  $l_\perp^\pm = x_\perp v_2 \pm  x_3 v_3$,
$l^\pm_\perp \cdot l^\pm_\perp = |l_\perp|^2$. Recall that
$l_\perp^2$ is fixed from the on-shell condition
Eq.(\ref{eq23})
and therefore
$x_3$ is expressed through $x_\perp$, $x_3 = \sqrt{|l_\perp|^2 - x_\perp^2}$.
We calculate $b_\pm = b(l^\pm)$ and remove
$x_3^2$ in favor of $l_\perp^2$ and $x_\perp^2$ where possible. We obtain
\ba
b_{\pm } =
b_1+ b_2 x_\perp \pm  x_3 b_3
+ b_5 \left ( 2 x_\perp^2 - l_\perp^2 \right )
+ b_6 l_\perp^2 \pm  b_7 x_\perp x_3.
\ea
Taking the sum and the difference of $b_{\pm}$, we arrive at
\be
\frac{\left ( b_++ b_- \right )}{2} =
b_1 + b_2 x_\perp  + b_5 \left ( 2 x_\perp^2 - l_\perp^2 \right ) +
b_6 l_\perp^2,\;\;\;
\frac{\left (b_+- b_- \right )}{2x_3}  =  b_3  + b_7 x_\perp.
\label{eqbpm}
\ee
The right hand sides of these equations are
polynomials in $x_\perp$. Therefore, we can apply discrete Fourier
transform with respect to $x_\perp$ to find
coefficients $b_2, b_5, b_1^{\rm eff} = b_1 + b_6 l_\perp^2$
as well as $b_3,b_7$  in Eq.(\ref{eqbpm}).

To determine the remaining  coefficients, we take  the vector
$l_\perp$ to be in the $v_2-v_4$ plane.
We choose\footnote{One should not be confused by
the mismatch in dimensions. In an actual computation,
we always rescale all dimension-full variables by the center
of mass collision energy,
to turn them into order one numbers.}  $x_2 = \sqrt{l_\perp^2 -1}$
and consider three  different vectors
$l_\perp^{(a)}   = x_2 v_2 +  v_4$,
$l_\perp^{(b)}   = -x_2 v_2 +  v_4$,
$l_\perp^{(c)}   = -x_2 v_2  - v_4$.
We use the notation  $b_{\alpha} = b(xp + l_\perp^{(\alpha)})$
but  we assume that all terms with previously computed coefficients
are  subtracted when  $b(xp + l_\perp^{(\alpha)})$ is computed.
It is easy to see that the following relations hold
\ba
&& b_8 = \frac{\left ( b_a - b_b \right )}{2 \sqrt{l_\perp^2 -1}},\;\;\;
b_4 = \frac{\left ( b_a - b_c \right )}{2},\;\;\;\;
b_6 = -\frac{\left ( b_a - b_{1}^{\rm eff} - b_4
- b_8 \sqrt{l_\perp^2 -1} \right )}{3}.
\ea
Also, we find  $b_1 = b_1^{\rm eff} - l_\perp^2 b_6$ and, finally,
the term with $b_9$ can be obtained  by calculating the function
$b$ for  the momentum $l_\perp = v_3 + \sqrt{|l_\perp|^2 - 1}\; v_4 $.

We have just described a method to calculate  coefficients $b_{1..9}$
in a numerically stable way for arbitrary values of
$|l_\perp|$.  In the numerical program, we switch from
the discrete Fourier transform to the solution just described,
depending on the value of $|l_\perp|$.
However, the described  methods can only work {\it if}
the decomposition of the loop
momentum, as in Eq.(\ref{eq222}),  exists.  A glance at
Eq.(\ref{eq23}) makes it clear that the decomposition
fails for the {\it light-like}
momentum, $p^2 = 0$,  and we have to handle this case differently.
We describe a possible solution below.

Because we are interested in
one-loop calculations for infra-red
safe observables, it is
reasonable to assume that
the vector $p$ can  be  {\it exactly} light-like but
it is impossible for that vector to be {\it nearly} light-like,
since such kinematic configurations are, typically, rejected by
cuts\footnote{External particles with small masses are
obvious exceptions but rarely do we need to know observables for, say,
massive $b$-quarks in a situation when all kinematic invariants
are large.}.
Hence,  we have to modify the above analysis to allow for a
light-like external vector.  To this end, we choose a frame where
the four-vector in Eq.(\ref{eq22})  reads
$p = (E,0,0,E)$. We  introduce
a complimentary light-like vector $\bar p = (E,0,0,-E)$.   The loop
momentum is parameterized as  $l= x_1 p + x_2 \bar p + l_\perp$.
We denote the basis vectors of the transverse space as $v_{3,4}$;
they satisfy $v_i v_j = \delta_{ij},\; p_{} v_{3,4}  = 0,\; \bar p_{} v_{3,4}
= 0$.
The on-shell condition for the loop momentum fixes $x_2$
\be
x_2 = \frac{m_2^2 - m_1^2}{s},\;\;\; s = 2 p \bar p,
\label{eqlperp1}
\ee
and a linear combination of $x_1$ and $l_\perp^2$
\be
l_\perp^2 + m_2^2x_1 - m_1^2 (1+x_1) = 0.
\label{eqlperp}
\ee
The parameterization of the function ${\bar b}$ reads
\ba
&&  {\bar b}(l)  = b_1 + b_2 (\bar p l)  + b_3 (v_3 l) + b_4 (v_4 l)
+ b_5 (\bar p l) (\bar p l)
+ b_6 (\bar p l) (v_3 l)
\nonumber \\
&&\;\;\;\;\;\;\; +\; b_7 (\bar p l) (v_4 l)
+ b_8 ( (v_3 l)^2 - (v_4 l)^2 )
+ b_9 (v_3 l) (v_4 l).
\label{eq154}
\ea

We describe a procedure to find the coefficients
$b_1,..,b_9$ in a numerically stable way. To this end,
we  choose $x_1 = 0.5$. This fixes
$|l_\perp|^2$ and $x_2$ is fixed by the on-shell condition
Eq.(\ref{eqlperp1}).  The
freedom remains  to choose the {\it direction} of the vector $l_\perp$
in the $v_3-v_4$ plane. Consider  four different vectors
\be
l_\perp^{(a)} =  v_3 + x_4 v_4,\;
l_\perp^{(b)} = - v_3 + x_4 v_4,\;
l_\perp^{(c)} = v_3 - x_4 v_4,\;
l_\perp^{(d)} = -v_3 - x_4 v_4,
\label{eq30}
\ee
where $x_4 = \sqrt{l_\perp^2 -1 }$. We use vectors
$l^{(\alpha)} = x_1 p + x_2 \bar p + l_\perp^{(\alpha )}$, $\alpha = a,b,c,d,$
to calculate the  function $b^{(\alpha)} = b(l^{\alpha})$.
Using $b_a,...b_d$,  we can immediately find the coefficient
$b_9$
\be
b_9 = \frac{1}{4\sqrt{l_\perp^2 -1}}
\left ( b^{(a)} - b^{(c)} - b^{(b)} + b^{(d)} \right ).
\ee
For the determination of the remaining coefficients,
it is convenient to introduce two linear combinations
\ba
b_{47} =
\frac{1}{4\sqrt{l_\perp^2 -1}} \left ( b^{(a)} - b^{(c)} + b^{(b)} - b^{(d)} \right ),\;\;\;
b_{36} = \frac{1}{2} \left ( b^{(a)} - b^{(b)} \right )
- b_9 \sqrt{l_{\perp}^2 -1 }.
\label{eq09}
\ea

As the next step, we choose $x_1 = -0.5$. Note that this
changes the value of $l_\perp^2$ according to Eq.(\ref{eqlperp}).
We then repeat the calculation described above.
Our choices of momenta in the transverse plane  $l_\perp$
are the same
as in Eq.(\ref{eq30}) but, to avoid confusion, we emphasize that
$x_4$ has to be calculated with the new $|l_\perp|$. We will refer
to $b$ computed with those new vectors as $\tilde b^{(a)}$,
$\tilde b^{(b)}$, etc. We calculate $\tilde b_{47,36}$ by substituting
$b^{(\alpha)}  \to \tilde b^{\alpha}$ in
Eq.(\ref{eq09}).  It is easy to see that simple linear
combinations give the desired coefficients
\ba
b_4 = \frac{1}{2} \left ( b_{47} + \tilde b_{47} \right ),\;\;\;\;
b_7 = \frac{1}{s} \left ( b_{47} - \tilde b_{47} \right ),
\nonumber \\
b_3 = \frac{1}{2} \left ( b_{36} + \tilde b_{36} \right ),\;\;\;\;
b_6 = \frac{1}{s} \left ( b_{36} - \tilde b_{36} \right ).
\ea
Other coefficients, required
for the complete parameterization of the function $b(l)$
in Eq.(\ref{eq154}), are obtained along similar lines;
we do not discuss this further. However,
we emphasize that the procedure that we just
described is important for the computation
 of one-loop virtual amplitudes in a situation where both massless and massive
particles are involved.  In particular, it is heavily used in the
computation of NLO QCD corrections to top quark pair production
discussed in Ref.~\cite{ms} and in this paper.

As a final remark, we note that
there is another important   difference between reducing the two-point
function  to scalar integrals for a light-like
and a regular vector $p$. Indeed,
the two-point function
with $p^2 \ne 0$ can be immediately simplified
\be
\int \frac{{\rm d}^Dl}{(2\pi)^D} \frac{\bar{b}(l)}{d_1 d_2}
= b_1 \int \frac{{\rm d}^Dl}{(2\pi)^D} \frac{1}{d_1 d_2}.
\ee
This feature follows from  the fact that all but the $l$-independent
terms   in the function $b(l)$,  Eq.(\ref{eq145}),
vanish when integrated over the
directions of  the transverse space, cf.  Eq.(\ref{eqm123}).
Hence, the  only integral we need to know
is the  scalar two-point function. However, in case of
a light-like vector $p^2=0$, {\it three} master integrals
survive even after  averaging over the directions of the vector $l$ in
the (two-dimensional) transverse space
\be
\int \frac{{\rm d}^Dl}{(2\pi)^D} \frac{\bar{b}(l)}{d_1 d_2}
= \int \frac{{\rm d}^Dl}{(2\pi)^D} \frac{b_1 + b_2 (\bar p l)
+ b_5 (\bar p l)^2 }{d_1 d_2}.
\ee
Those integrals must be  included to the basis  of master integrals
in case when double cuts with a light-like vector are considered.
The calculation of those additional master integrals is straightforward;
for completeness, we give the results below for the equal mass
case $m_1 = m_2 = m$. We introduce
$d_1 = (l+k_1)^2 - m^2$, $d_2=(l+k_2)^2 - m^2$,
$k_2 - k_1 = p$, $p^2 = 0$, $\bar p k_1 = r_1$, $\bar p k_2 = r_2$,
$c_\Gamma = (4\pi)^{\epsilon-2}
\Gamma(1+\epsilon) \Gamma(1-\epsilon)^2/\Gamma(1-2\epsilon)$
and find ($D = 4-2\epsilon$)
\ba
&& \frac{\mu^{2\epsilon}}{\mathrm{i} c_\Gamma}
\int \frac{{\rm d}^Dl }{(2\pi)^D}\;\;
\frac{l \bar p}{d_1 d_2}
=
\frac{\left ( r_1 + r_2\right )}{2}\;
\left ( -\frac{1}{\epsilon} + \ln \left( \frac{m^2}{\mu^2} \right) \right ),
 \\
&& \frac{\mu^{2\epsilon}}{\mathrm{i}c_\Gamma}
\int \frac{{\rm d}^Dl }{(2\pi)^D}\;\;
\frac{(l \bar p)(l \bar p)}{d_1 d_2}
=
\frac{(r_1^2 + r_1 r_2  + r_2^2)}{3}\;
\left ( -\frac{1}{\epsilon} + \ln \left( \frac{m^2}{\mu^2} \right)\right ).
\ea

\section{Technical details  of the calculation}
\label{sect3}

In this Section, we collect all the information pertinent to
the technical
details of the computation. As with  any next-to-leading order
calculation, we need to deal with one-loop virtual corrections,
real-emission corrections  and the soft-collinear
subtraction terms.
Each of these contributions requires a different treatment and
we discuss them in turn. Much of what we say in this Section
can be found in the literature but we decided to summarize all the
required details in this paper, for completeness.

\subsection{The decay of the top quark}

As we pointed out in the Introduction, we include decays of the
top quark, albeit in leading order in perturbative QCD.
The way this is done can be illustrated by considering tree-level
process; NLO QCD  effects do not introduce additional subtleties.
Our discussion follows Ref.~\cite{ms}.
Consider production of a $t \bar t$  pair and a jet in the collision
of partons $a$ and $b$.  Top quarks decay into bottom quarks
and $W$-bosons.  For definiteness we consider
leptonic decays of the $W$ bosons
  $ab \to t \bar t j \to (\bar b l^- \bar \nu)
(b l^+ \nu)j$
but there is not much difference with hadronic decays as long as we restrict
ourselves to leading order in top decays.
The amplitude for this process is written as
\be
A = \bar \Psi_\alpha(t^* \to b l^+ \nu)\;
A_{\alpha \beta} (ab \to t^* \bar t^* + j) \;
\Xi_\beta ( \bar t^* \to \bar b l^- \bar \nu),
\label{eq44}
\ee
where $\Psi$ and $\Xi$ are the off-shell fermion currents
and $\alpha,\beta$ are the Dirac indices.
The fermion currents contain $t$($\bar t$) propagators; for example
\be
\bar \Psi_\alpha(t^* \to b l^+ \nu) =
{\tilde A}_\gamma(t^* \to b l^+ \nu)
\frac{\mathrm{i} (\Fslash{p}_t+m_t)_{\gamma \alpha}}{p_t^2 - m_t^2 + \mathrm{i}m_t \Gamma_t},
\label{eq4444}
\ee
where ${\tilde A}_\gamma$ is  a sub-amplitude that describes  transition
of an off-shell fermion $t^*$ to a final state $b l^+ \nu$.

We note that radiation of a jet off the top quark decay products
is not included in Eq.(\ref{eq44}).  Therefore, as written,
Eq.(\ref{eq44})  is   {\it not gauge invariant}. However, this problem
disappears once we take the on-shell limit for both $t$ and $\bar t$.
To take the on-shell limit, we imagine squaring the amplitude
$A$ and integrating over the phase-space of the final state particles.
In the limit $\Gamma_t/m_t \to 0$, propagators that appear in Eq.(\ref{eq4444},\ref{eq44})
become proportional to the delta-function
\be
\frac{1}{(p_t^2 - m_t^2)^2 + m_t^2 \Gamma_t^2}
= {\Bigg |}_{\Gamma_t/m_t \to 0} =
\quad \quad \frac{2\pi}{2m_t\Gamma_t} \; \delta (p_t^2-m_t^2).
\label{eq456}
\ee
Finally, we factorize  the phase-space into a phase-space for
$t \bar t  j$  and a  phase-space for top decay products and observe
that  the $\delta$-function in Eq.(\ref{eq456}) forces top quarks
on their mass-shells. This separates production and decay stages and
ensures that those stages are separately gauge invariant.
The top decays are then conveniently
implemented by making the substitution in
Eq.(\ref{eq44})
\ba
&& \bar \Psi_\alpha(t \to b l^+ \nu) \to \bar U_\alpha (p_t)
= {\tilde A}_\gamma(t \to b l^+ \nu)
 \frac{\mathrm{i}(\Fslash{p}_t + m_t)_{\gamma \alpha}}{\sqrt{2m_t \Gamma_t}},
\nn \\
&&
\Xi_\alpha ( \bar t \to \bar b l^- \bar \nu) \to
V_\alpha(p_\tb) = \frac{\mathrm{i} (-\Fslash{p}_\tb + m_t)_{\alpha \gamma}}{\sqrt{2m_t \Gamma_t}}
{\tilde A}_\gamma(\bar t \to \bar b l^- \bar \nu).
\ea
We emphasize that $U$ and $V$ are conventional spinors with
 definite polarizations that are fully determined by momenta and helicities of
the top quark decay products.  In fact, direct multiplication of
spinors and Dirac matrices is an efficient way to reconstruct
$U$ and $V$ in a numerical program for each phase-space point. Finally,
we note that similar considerations work for one-loop corrections
and the subtraction terms.

\subsection{One-loop amplitudes}
Our calculation of the one-loop amplitudes
is based on the method of generalized
$D$-dimensional unitarity. In this Section, we describe the
building blocks that enter  the calculation.

The one-loop amplitudes, that we compute in this paper,
require renormalization of the bare top quark mass
$m_{t,{\rm bare}} = Z_m m_t$  , the bare top quark
field $\psi_{t,\rm bare} = \sqrt{Z_2} \psi_t $
and the strong coupling constant $g_s$.
For massive quarks, the mass renormalization
and the wave function renormalization
are performed on-shell.  The renormalization constants
depend on the regularization scheme; in this paper all the results
for one-loop amplitudes are given
in the four-dimensional helicity (FDH) scheme,  introduced
in Refs.~\cite{fdh1,fdh2}. In that  scheme, the on-shell
renormalization constants read \cite{egkm}
\be
Z_m = Z_2 = 1 - C_{\mathrm{F}} g_s^2 c_\Gamma
\left ( \frac{\mu^2}{m_t^2} \right )^\epsilon
\left ( \frac{3}{\epsilon} + 5 \right ) + {\cal O}(g_s^4,\epsilon),
\ee
where $C_{\mathrm{F}} = (N_c^2-1)/(2N_c)$, $N_c = 3$,
is the conventional QCD color factor.  Finally, we note that
the renormalization of the coupling constant is the same as in the
${\overline {\rm MS}}$-scheme but a finite shift needs to be applied
\cite{Kunszt:1993sd}
to relate $\alpha_s^{\rm FDH}$ and  conventional $\alpha_s^{\overline {\rm MS}}$
\be
\alpha_s^{\rm FDH} =
\alpha_s^{\overline {\rm MS}}
\left ( 1 + \frac{N_c}{12} \frac{\alpha_s^{\overline {\rm MS}}}{\pi}
\right ).
\ee

Objects that we compute within the unitarity framework are
the so-called primitive amplitudes \cite{bdk}. Regular amplitudes and
primitive amplitudes are related through the color decomposition
procedure. Below we give the decomposition of all amplitudes required
for our computation.
There are two types of tree- and one-loop amplitudes
that need to be considered to calculate  production of a top quark
pair and a jet  in hadron collisions:
$0 \to \bar t t + ggg$ and $ 0 \to \bar t t + q \bar q + g$.
We first consider the two-quark $(n-2)$-gluon process.
At tree-level,
the $0 \to \bar t + t + (n-2)~{\rm gluons}$ scattering amplitude
reads
\be
{\cal A}_n^{\rm tree} = g_s^{n-2}
\sum \limits_{\sigma \in S_{n-2}}^{} \left ( T^{a_{\sigma(3)}}
...T^{a_{\sigma(n)}} \right )_{i_2}^{{\bar i}_1}
A_n^{\rm tree}\left( 1_{\bar t}, 2_t,\sigma(3)_g,...,\sigma(n)_g \right ),
\label{ttng_tree}
\ee
where $A_n^{\rm tree}$ is the tree (left) primitive amplitude and
$S_{n-2}$ is the permutation group of $(n-2)$ elements.  The
${\rm SU}(3)$ generators are normalized as ${\rm Tr}(T^a T^b) = \delta^{ab}$
and satisfy the commutation relations
\be
[T^a, T^b] = -F^c_{ab} T^c.
\ee
This normalization allows us to use color-stripped Feynman
rules \cite{Bern:1994fz,Bern:1997sc,rules} to compute primitive amplitudes.


A similar expression for the one-loop amplitude is more complicated.
Using the color basis of  Ref.~\cite{duca}, we can write the one-loop
amplitude as a linear combination of left primitive amplitudes
\ba
{\cal A}_{n}^{\rm 1-loop}
= && g_s^n
\sum \limits_{p=2}^{n} \sum \limits_{\sigma \in S_{n-2}}^{}
\left ( T^{x_2} T^{a_{\sigma_3}}...T^{a_{\sigma_p}} T^{x_1}
\right )_{i_2}^{{\bar i}_1}
\left ( F^{a_{\sigma_{p+1}}}...F^{a_{\sigma_n}} \right )_{x_1 x_2}
\nonumber \\
&&
\times
(-1)^n A_L^{[1]} \left (1_{\bar t},\sigma(p)_g,...,\sigma(3)_g,2_t,\sigma(n)_g,...,
\sigma(p+1)_g \right )
\nonumber \\
&& + \frac{1}{N_c}
\sum \limits_{j=1}^{n-1} \sum \limits_{\sigma \in S_{n-2}/S_{n;j}}
{\rm Gr}^{(q \bar q)}_{n;j} \left (\sigma_3,...,\sigma_n \right )
\Bigg ( n_f
A_{L;j}^{[1/2,q]} \left( 1_{\bar t},2_t,\sigma(3)_g,..,\sigma(n)_g \right )
\nonumber \\
&& +
A_{L;j}^{[1/2,t]} \left( 1_{\bar t},2_t,\sigma(3)_g,...,\sigma(n)_g \right )
\Bigg ),
\label{eq985}
\ea
where $n_f=5$ is the number of quark flavors that are considered massless.
In Eq.(\ref{eq985}),
for  $p = 2$, the factor
$(T...T)^{\bar i_1}_{i_2} \to (T^{x_2} T^{x_1} )_{i_2}^{{\bar i}_1}$
and for $p=n$ the factor $(F...F)_{x_1 x_2} \to \delta_{x_1 x_2}$.
In the second term in Eq.(\ref{eq985}),
$S_{n;j}$ is the subgroup of $S_{n-2}$ which
leaves ${\rm Gr}^{(\bar q q)}_{n;j}$ invariant.  Primitive
amplitudes $A_{L;j}^{[1/2,q]}$, $A_{L;j}^{[1/2,t]}$ contain
one closed loop of massless and massive (top quarks) fermions, respectively.
The color factors read
\ba
&& {\rm Gr}_{n;1}^{(\bar q q)}(3,...,n) = N_c \left ( T^{a_3}.
..T^{a_n} \right ),\;\;
{\rm Gr}_{n;2}^{(\bar q q)}(3;4,...,n) = 0,\;\;\;
{\rm Gr}_{n;n-1}^{(\bar q q)}(3,...,n) =
{\rm Tr} \left ( T^{a_3}...T^{a_n} \right ) \delta_{i_2}^{{\bar i}_1},
\nonumber \\
&& {\rm Gr}^{(\bar q q)}_{n;j}\left (3,...,j+1;j+2,...,n \right )
= {\rm Tr} ( T^{a_3}...T^{a_{j+1}})(T^{a_{j+2}}...T^{a_n})_{i_2}^{{\bar i}_1},\;\;
j=3,...,n-2.
\ea

\begin{figure}[t]
\begin{center}
\epsfig{file=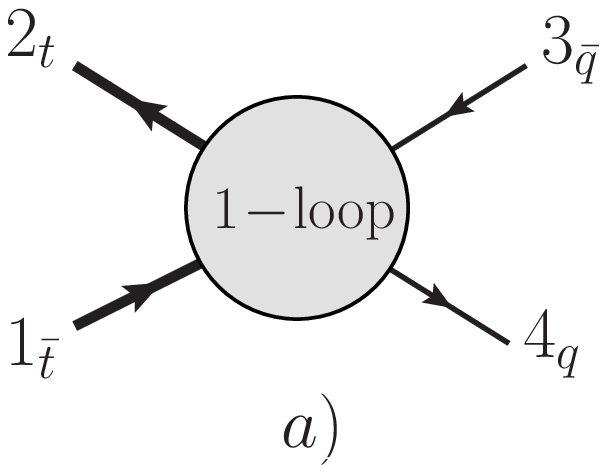, angle=0,width=0.25\textwidth}
\hspace{1cm}
\epsfig{file=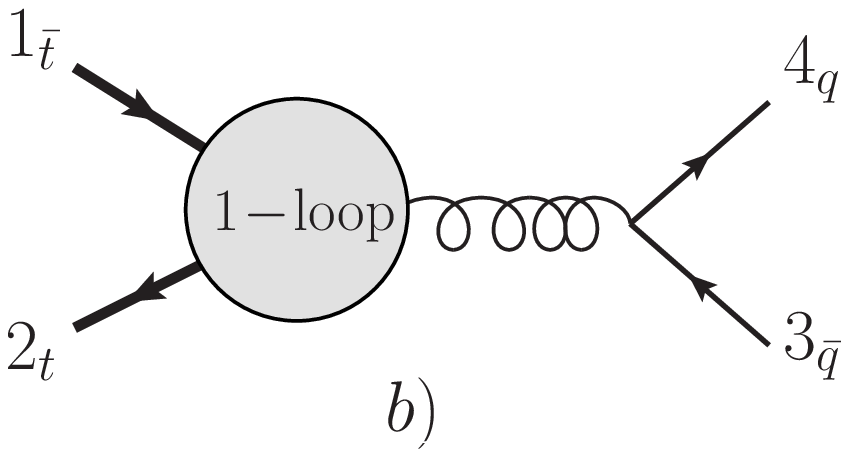, angle=0,width=0.35\textwidth}
\\[0.3cm]
\epsfig{file=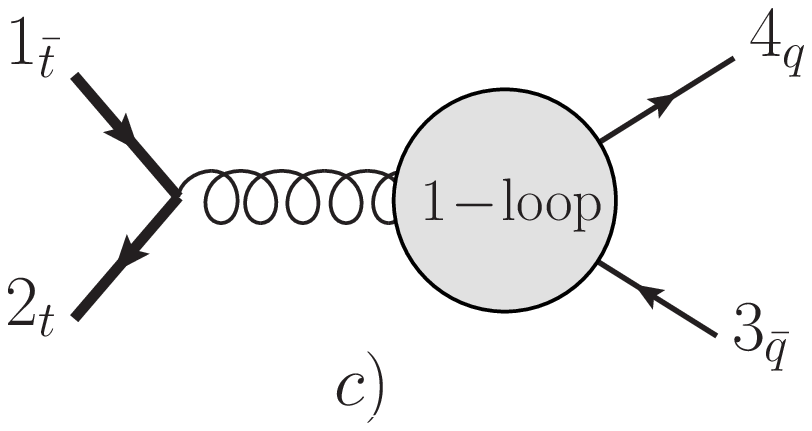, angle=0,width=0.35\textwidth}
\hspace{1cm}
\epsfig{file=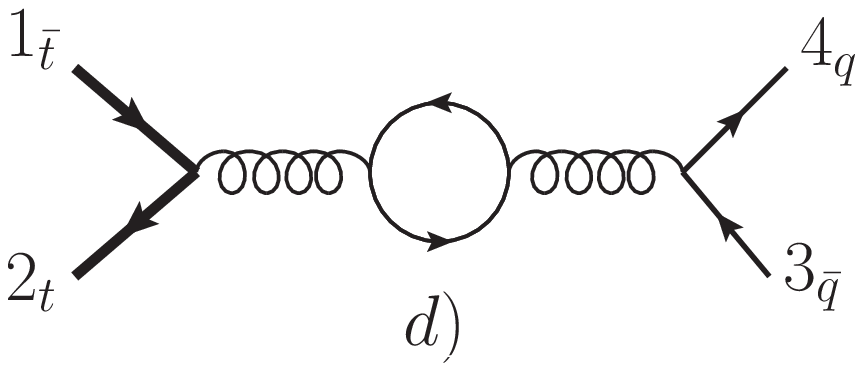, angle=0,width=0.4\textwidth}
\caption{Prototype parent
diagrams for primitive amplitudes with four quarks,  and a
gluon for classes ``a'', ``b'', ``c'' and closed fermion loops (d).  The gluon can be inserted
in four possible ways into any of the prototype graphs, leading to
four primitive amplitudes in each case.}
\label{fig3}
\end{center}
\end{figure}

As the next step,
we describe the color decomposition of
$0 \to \bar t t + \bar q q +g $
amplitude \cite{egkmz}, where $q$ is a massless quark.
The color decompositions of the tree and the one-loop amplitudes read
\begin{eqnarray}
 \B^\tree (1_{\bar t},2_t,3_\qb,4_q,5_g) & =&   g_s^3 \Bigg[
(T^{a_5})_{i_4}^{~\ib_1} \delta_{i_2}^{~\ib_3} B^\tree_{1}
 +\frac{1}{N_c}
(T^{a_5})_{i_2}^{~\ib_1} \delta_{i_4}^{~\ib_3}
B^\tree _{2}
\nn \\
&&
+ (T^{a_5})_{i_2}^{~\ib_3} \delta_{i_4}^{~\ib_1}
B^\tree _{3}
    +\frac{1}{N_c} (T^{a_5})_{i_4}^{~\ib_3}
\delta_{i_2}^{~\ib_1}B^\tree _{4}\Bigg];\\
\B^\oneloop (1_{\bar t},2_t,3_{\bar q},4_q,5_g) &=&  g_s^5 \Bigg[
N_c (T^{a_5})_{i_4}^{~\ib_1} \delta_{i_2}^{~\ib_3} B_{1}
 +(T^{a_5})_{i_2}^{~\ib_1} \delta_{i_4}^{~\ib_3}
B_{2}
\nn \\
&+&
N_c (T^{a_5})_{i_2}^{~\ib_3} \delta_{i_4}^{~\ib_1}
B_{3}
    +(T^{a_5})_{i_4}^{~\ib_3}
\delta_{i_2}^{~\ib_1}B_{4}\Bigg].
\end{eqnarray}
In each of these one-loop color-ordered amplitudes we separate
terms with the  closed massless fermion loop, with the top quark
loop and all other contributions
\be
B_{i} = B^{[1]}_{i}+ \frac{n_f}{N_c} B^{[1/2,q]}_{i}+
\frac{1}{N_c} B^{[1/2,t]}_{i}, \qquad i=1,2,3,4.
\ee
The amplitudes $B^{[1]}_{i}$ and $B^{[1/2,q(t)]}_{i}$ can be written as
linear combinations of primitive amplitudes.  Those primitive
amplitudes are illustrated  in Fig.~\ref{fig3}. Explicit
expressions  for
color-ordered amplitudes
$B^{[1]}_{i}$ and $B^{[1/2,q(t)]}_{i}$ in terms
of the primitive amplitudes are given in Ref.\cite{egkmz}; they can be found
in the Appendix~A.

In Ref.~\cite{Dittmaier:2008uj}, numerical results for spin-averaged
squared
amplitudes  for the virtual corrections were given for
a particular kinematic point. In Appendix~B, we give numerical results
for most of the helicity amplitudes for that kinematic point.
We have checked the  calculation in several ways.  For example, our
primitive amplitudes  have correct $1/\epsilon$ poles \cite{Catani:2000ef}
and are gauge invariant, once mass counter-terms
are accounted for.  We  performed
a diagrammatic check comparing amplitudes computed through unitarity
and on-shell matrix elements with a similar, but independent, implementation
of the OPP procedure \cite{opp}
which is applicable to individual Feynman diagrams.
Finally, we have checked that when we compute the spin-averaged amplitude
squared for the kinematic point considered in Ref.~\cite{Dittmaier:2008uj},
we find full agreement with their results.

A well-known  problem that occurs in the process of the reduction of
one-loop diagrams to any basis of scalar integrals is the appearance
of unphysical singularities\footnote{Those unphysical singularities
are caused by the so-called Gram determinants.}
at the intermediate stages of the calculation.
Those singularities  cancel in the final result as a matter of principle
but it has been proven to be difficult to achieve such cancellations in practice
if reductions are performed numerically. Within the current calculation,
we control potential numerical instabilities by requiring that the
$1/\epsilon$ pole of a primitive amplitudes
is calculated correctly to
within ${\cal O}(10^{-4})$. If this condition is not fulfilled,
the program switches from double to quadruple precision and re-calculates the
offending amplitude.  The number of points that needs to be
re-calculated is typically small, below a percent for partonic
 channels that we consider.

\subsection{Real emission processes and soft-collinear
subtraction counter-terms}

Computation of the real emission corrections and subtraction counter-terms
is an important part of any NLO computation.
There are three generic processes that
need to be considered:  $ 0 \to \bar t t + gggg$,
$0 \to \bar t t + \bar q q + gg$ and $0 \to \bar t t + \bar q q + \bar q' q'$.
Once amplitudes for those processes are known,
all partonic channels that contribute to the real emission correction
to the hadronic production of $t \bar t + {\rm jet}$ can be  constructed.
We employ color decomposition to express those amplitudes through color-ordered
ones and use Berends-Giele recurrence relations \cite{bg}
 to compute the latter.
Below results for  color decomposition of the relevant tree  amplitudes
are summarized.

The color decomposition for the amplitude
$0 \to \bar t t + gggg$ has already been
given in Eq.(\ref{ttng_tree}). The color decomposition for the amplitude
$0 \to \bar t t + \bar q q + gg$   in double index notation for the
gluon field  reads \cite{msz}
\ba
&& {\cal A}(\bar t, t, \bar q, q, g_1,g_2)=  g_s^4
\sum \limits_{\sigma \in S_2}^{} \Bigg (
\delta^{i_t}_{j_\qb} \delta^{i_q}_{j_{\sigma(1)}} \delta^{i_{\sigma(1)}}_{j_{\sigma(2)}}
\delta^{i_{\sigma(2)}}_{j_\tb} A(\bar t, t,\qb,q,g_{\sigma(1)},g_{\sigma(2)})
\nonumber  \\
&&
+\delta^{i_t}_{j_{\sigma(1)}} \delta^{i_{\sigma(1)}}_{j_{\sigma(2)}} \delta^{i_{\sigma(2)}}_{j_\qb}
\delta^{i_q}_{j_\tb} A(\bar t, t,g_{\sigma(1)},g_{\sigma(2)},\qb,q)
+\delta^{i_t}_{j_{\sigma(1)}} \delta^{i_{\sigma(1)}}_{j_\qb} \delta^{i_q}_{j_{\sigma(2)}}
\delta^{i_{\sigma(2)}}_{j_\tb} A(\bar t, t,g_{\sigma(1)},\qb,q,g_{\sigma(2)})
\nonumber \\
&& - \frac{1}{N_c}
\delta^{i_t}_{j_{\sigma(2)}} \delta^{i_{\sigma(2)}}_{j_{\sigma(1)}}
\delta^{i_{\sigma(1)}}_{j_{\bar t}}
 \delta^{i_q}_{j_\qb}  A(\bar t,g_{\sigma(1)},g_{\sigma(2)},t,\bar q,  q)
- \frac{1}{N_c}\delta^{i_t}_{j_{\sigma(1)}} \delta^{i_{\sigma(1)}}_{j_\tb} \delta^{i_{q}}_{j_{\sigma(2)}}
 \delta^{i_{\sigma(2)}}_{j_\qb}  A(\bar t,g_{\sigma(1)}, t, \bar q, g_{\sigma(2)},q)
\nonumber \\
&& - \frac{1}{N_c}
\delta^{i_q}_{j_{\sigma(2)}} \delta^{i_{\sigma(2)}}_{j_{\sigma(1)}}
\delta^{i_{\sigma(1)}}_{j_{\bar q}}
 \delta^{i_t}_{j_{\bar t}}  A(\bar t, t,\bar q, g_{\sigma(1)},g_{\sigma(2)},q)
\Bigg ). \nn
\ea
Finally, the color decomposition for the six quark amplitude for
non-identical quarks
$0 \to \bar t t + \bar q q + \bar f f$ reads
\ba
{\cal A}(\bar t, t, \bar q, q , \bar f, f) &=&
\delta^{i_t}_{j_\qb} \delta^{i_q}_{j_{\bar f}} \delta^{i_f}_{j_{\bar t}}
 A(\bar t,t, \bar q, q, \bar f, f)
+ \delta^{i_t}_{j_{\bar f}} \delta^{i_f}_{j_\qb} \delta^{i_q}_{j_{\bar t}}
A(\bar t,t, \bar f, f, \bar q, q)
\nonumber \\
&& -\frac{1}{N_c} \left (
 \delta^{i_q}_{j_\qb} \delta^{i_t}_{j_{\bar f}} \delta^{i_f}_{j_{\bar t}}
 + \delta^{i_t}_{j_{\bar t}} \delta^{i_f}_{j_\qb} \delta^{i_q}_{j_{\bar f}}
  -\frac{1}{N_c}
\delta^{i_t}_{j_{\bar t}} \delta^{i_f}_{j_{\bar f}} \delta^{i_q}_{j_\qb}
\right )A(\bar t, f, \bar  q, q, \bar f, t)
\nonumber \\
&& -\frac{1}{N_c}\left (
\delta^{i_f}_{j_{\bar f}} \delta^{i_t}_{j_\qb} \delta^{i_q}_{j_{\bar t}}
+ \delta^{i_q}_{j_\qb} \delta^{i_f}_{j_{\bar t}} \delta^{i_t}_{j_{\bar f}}
  -\frac{1}{N_c}
\delta^{i_t}_{j_{\bar t}} \delta^{i_f}_{j_{\bar f}} \delta^{i_q}_{j_\qb} \right )
 A(\bar t, f, \bar f, t,\bar q,q)
\nonumber \\
&& - \frac{1}{N_c} \left (
\delta^{i_f}_{j_{\bar f}} \delta^{i_t}_{j_\qb} \delta^{i_q}_{j_{\bar t}}
+ \delta^{i_t}_{j_{\bar t}} \delta^{i_f}_{j_\qb} \delta^{i_q}_{j_{\bar f}}
  -\frac{1}{N_c} \delta^{i_t}_{j_{\bar t}}
 \delta^{i_f}_{j_{\bar f}} \delta^{i_q}_{j_\qb} \right )
A(\bar t, q, \bar f, f, \bar q, t).
\ea

The amplitude for the identical quarks
${\cal A}(\bar t, t, \bar q, q , \bar q, q)$ can be obtained from the
previous amplitude by the anti-symmetrization procedure
\be
{\cal A}(\bar t, t, \bar q, q , \bar q, q)
=
{\cal A}(\bar t, t, \bar q, q , \bar f, f)
-
{\cal A}(\bar t, t, \bar q, f , \bar f, q).
\ee

We note that many of the  amplitudes
required for  the computation of the real emission corrections
are also employed for the computation of the virtual
corrections within the unitarity framework. We use one and the
same program that computes the on-shell tree  amplitudes for
complex or real external momenta,
as needed for the virtual and the real corrections, respectively.
  We checked our calculation
of the real emission matrix elements against Madgraph \cite{Stelzer:1994ta}
for all the partonic  channels.

As it is typical
for NLO calculations, when we integrate the real emission matrix
elements with $t \bar t$ and two additional partons in the final
state over  the phase-space constrained by the requirement that
a $t \bar t$ pair and at least one jet is observed,
we encounter  infra-red and collinear divergences.
Those divergences
are removed by the subtraction procedure. We use dipole formalism of
Ref.~\cite{Catani:1996vz} extended  to deal with the
QCD radiation off  massive particles in Ref.~\cite{Catani:2002hc}.
We note that in the original publications, the subtraction terms were
integrated over full unresolved phase-space which is not very convenient.
An optimization of the subtraction procedure, where subtractions
are only performed if the event kinematics is close to singular,
was suggested in Ref.~\cite{nagy}. Dipoles  integrated over the
restricted phase-space including the case of massive emitters and spectators
can be found
in  Ref.~\cite{Bevilacqua:2009zn,mcfm,Campbell:2004ch,Campbell:2005bb}.
In the actual implementation of the subtraction procedure, we closely
follow Ref.~\cite{mcfm}.  We have checked that our results do not
depend on the parameter that restrict the integration over the dipole
phase-space; this is a useful way to check the consistency of the
implementation of the subtraction terms.

\begin{figure}[t]
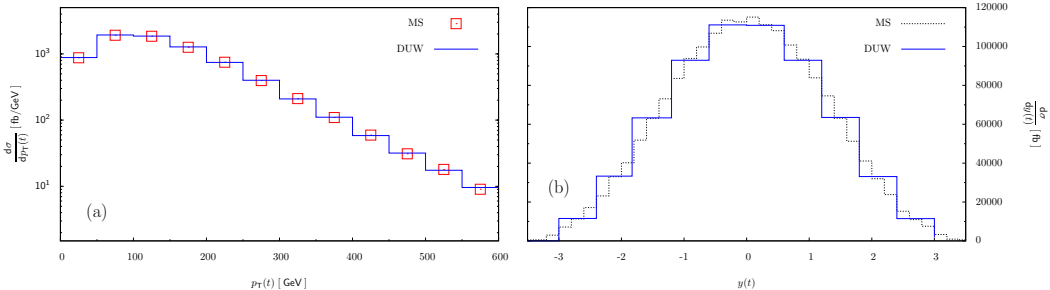

\begin{center}
  \scalebox{0.4}{\input{LHC_DUW_Fig01.tex}}
  \scalebox{0.4}{\input{LHC_DUW_Fig02.tex}}
\caption{Top quark transverse momentum (a)  and rapidity (b)
distributions (MS) computed through NLO QCD for
$pp \to t \bar t j$,  at the LHC ($\sqrt{s} = 14~{\rm TeV}$)
compared to the results of
Refs.~\cite{Dittmaier:2007wz,Dittmaier:2008uj} (DUW).  The renormalization and
factorization scales are set to $\mu = m_t$.}
\label{fig4}
\end{center}
\end{figure}

\section{Results}
\label{sectr}

In this Section we present the results of the  calculation of the
NLO QCD
corrections to $t \bar t$ pair production in association with one hard
jet at the Tevatron and the LHC.
 We begin by comparing our results with
that of Refs.~\cite{Dittmaier:2007wz,Dittmaier:2008uj}.
Then we present  the   results that include the top quark decays.

\subsection{Input parameters and comparison with existing results}

For the comparison with Refs.~\cite{Dittmaier:2007wz,Dittmaier:2008uj},
we need  to choose identical input parameters.
To this end, we use the
top quark mass $m_t = 174~{\rm GeV}$.
Also, the parton distribution functions CTEQ6L1/6M
have been updated since the results of
Ref.~\cite{Dittmaier:2007wz,Dittmaier:2008uj} appeared; therefore, for the
comparison we employ their older versions.
Finally, the $k_\perp$-clustering
algorithm with
$R_{ij}  = \sqrt{(y_i-y_j)^2 - (\phi_i - \phi_j)^2} = 1$
and  $E_\perp$-weighted  recombination scheme \cite{ellissoper}
is used to define jets in  Refs.~\cite{Dittmaier:2007wz,Dittmaier:2008uj}.
The jet transverse momentum  cut $p_{\perp,j}
= 50(20)~{\rm GeV}$ is employed
for the LHC (Tevatron), respectively.
When we use those parameters
in  the calculation,
we obtain good agreement with Refs.~\cite{Dittmaier:2007wz,Dittmaier:2008uj}.
We quote our results for the LHC ($\sqrt{s} = 14~{\rm TeV}$),
with  the factorization and the renormalization scales set to $m_t$.
For the total cross-section we obtain
\be
\sigma^{\rm NLO}(pp \to t \bar t j) = 375.8(1.0)~{\rm pb},
\ee
which agrees well with the result $376.2~{\rm pb}$ quoted in
Refs.~\cite{Dittmaier:2007wz,Dittmaier:2008uj}.
We note that the NLO QCD $pp \to t \bar t j$ cross-section at the LHC for the input parameters
of Refs.~\cite{Dittmaier:2007wz,Dittmaier:2008uj} was also reported in Ref.~\cite{Bevilacqua:2010ve} where the result $376.6~{\rm pb}$ was found.
We have checked that
similar level of agreement -- one percent or better --
persists for cross-sections evaluated for other values of the renormalization
and factorization scales, for both the Tevatron and the LHC.  We have
also verified that we reproduce kinematic distributions
presented in Refs.~\cite{Dittmaier:2007wz,Dittmaier:2008uj}.
As an illustration,  our results
for the transverse momentum and the rapidity distributions
of the top quarks at the LHC, computed through NLO QCD,
are compared with similar distributions from
Refs.~\cite{Dittmaier:2007wz,Dittmaier:2008uj}
in Fig.~\ref{fig4}. Good agreement between the two results
is apparent. We also confirm
an observation in  Refs.~\cite{Dittmaier:2007wz,Dittmaier:2008uj}
that the forward-backward asymmetry of top quarks
in $t \bar t +{\rm jet}$ production at the Tevatron
is  significantly
reduced if NLO QCD effects are taken into account. For $\mu = m_t$
and the jet transverse momentum cut of $20~{\rm GeV}$,
we find the  forward-backward asymmetry computed through NLO QCD
to be   $-2.28\%$, in agreement with $-2.27\%$ reported in
Refs.~\cite{Dittmaier:2007wz,Dittmaier:2008uj}. We
discuss the forward-backward asymmetry in more detail below, when
we report on our calculation that includes decays of top quarks.

\begin{figure}[!t]
 \begin{center}
  \scalebox{0.4}{\input{TEV12_Fig01.tex}}
  \scalebox{0.4}{\input{TEV12_Fig02.tex}} \\[8mm]
  \scalebox{0.4}{\input{TEV12_Fig04.tex}}
  \scalebox{0.4}{\input{TEV12_Fig05.tex}} \\[8mm]
  \scalebox{0.4}{\input{TEV12_Fig06.tex}}
  \scalebox{0.4}{\input{TEV12_Fig07.tex}}
\end{center}
\caption{ Various distributions for the process
$p \bar p \to (t \to  l^+ \nu b)
+ (\bar t \to j_1 j_2 \bar b) + j
$
at the Tevatron at leading (blue)
and next-to-leading order (red) in perturbative QCD.
The bands correspond to the choice of the renormalization
and factorization scales $\mu = [m_t/2, m_t, 2m_t]$.
We show distributions of the transverse momentum (a) and  rapidity of the
positron (b), the total transverse energy $H_\perp$ (c),
the transverse momentum (d)
and the rapidity (e) distributions of
the fifth hardest jet and the missing transverse momentum (f).}
\label{fig5}
\end{figure}

Having performed an  extensive comparison with results
of Refs.~\cite{Dittmaier:2007wz,Dittmaier:2008uj},
we turn to the discussion of observables that can be studied
if decays of top quarks are accounted for in the calculation.
We consider two primary scenarios -- $t \bar t j$ as the
signal process and as the background process to weak boson fusion production
of the Higgs boson.  It is important to consider  these different
scenarios  because  we want to understand to what extent
applied cuts affect the
radiative corrections.
We begin with the discussion of the $pp(p\bar p) \to t \bar t j$ production
in the kinematic region which is usually employed
to study top quarks. We perform
separate studies for the Tevatron and the LHC.
We choose two   center of mass energies $\sqrt{s} = 7~{\rm TeV}$
and $\sqrt{s} = 14~{\rm TeV}$  for the LHC
and   one center of mass energy $\sqrt{s} =
1.96~{\rm TeV}$ for the Tevatron.  We use the following values
for the top quark and the $W$-boson masses
$m_t = 172~{\rm GeV}$, $m_W = 80.419~{\rm GeV}$ and   employ
CTEQ6L1 and CTEQ6.1M parton distribution functions in
LO and NLO calculations,
respectively. We employ  the $k_\perp$ jet algorithm with $R_{ij} = 0.5$
and the  four-momentum recombination scheme\footnote{
We note that  $R_{ij} = 0.5$ in the $k_\perp$
algorithm roughly corresponds to $R_{ij} = 0.4$ in a typical
cone algorithm \cite{dgms}.}.
The couplings of the $W$-boson to fermions are obtained from
the Fermi constant $G_\mathrm{F} = 1.16639 \cdot 10^{-5} \,{\rm GeV}^{-2}$.
We emphasize that  results reported in this paper are calculated
with  {\it on-shell}
top quark decays at leading order in QCD; radiative corrections
to the decays are {\it not} included.
We take  the CKM matrix to be the identity  matrix.
We consider  the $W$-boson
in $t \to Wb$ decay to be on the mass shell.
With these approximations,
we obtain the  leading order top
quark decay width
 $\Gamma_t = 1.47~{\rm GeV}$.
We set the width of the $W$ boson
to $2.14~{\rm GeV}$. With our input parameters, we find the
following branching ratios
${\rm Br}(W^+ \to l^+ \nu) = 10.6~\%$,
${\rm Br}(W^+ \to {\rm hadrons}) = 64~\%$. We note
that the hadronic branching
fraction is slightly lower than the experimental value $67~\%$.

\subsection{$t \bar t + {\rm jet}$ production in lepton + jets channel at the Tevatron}

We begin our discussion with the Tevatron, $\sqrt{s} = 1.96~{\rm TeV}$.
 The measurement of
the $p \bar p \to t \bar t + {\rm jet}$
at the Tevatron was performed
by the CDF collaboration and there is a public note \cite{joey}
that describes the  details of the experimental setup and
the results of the measurements. For our purposes,
we can take all the selection
criteria for $p \bar p \to t \bar t j$
from that reference. We consider events where
top quark decays leptonically and the anti-top quark decays
hadronically. We require that there are five or more jets in the event
and that two of these jets are $b$-jets.
It is to be understood that a  $b$-jet
is {\it defined} as a jet that originates from the $tWb$ vertex.
For $p \bar p \to t \bar t j$ process, this leads
to a misidentification of $b$-jets, that originate from
$t \bar t$  production,
as non-$b$ jets. However,  this is a minor issue since $b$-jets
in the production process appear infrequently.
The lepton transverse momentum and the missing
energy in the event should satisfy $p_{\perp,l}> 20~{\rm GeV}$,
$E_\perp^{\rm miss} > 20~{\rm GeV}$. The jet
 transverse momenta   are required to be
larger than
$p_{\perp,j} > 20~{\rm GeV}$ and jets must be central $|y_j| < 2$.
There is an additional cut on the transverse energy in
the event $H_\perp = \sum_{j} p_{\perp,j} +
p_{\perp,e} + E_{\perp}^{\rm miss}$,  needed to better
discriminate against the background. It is required that
$H_\perp > 220~{\rm GeV}$.

We first present the results for total cross-sections. We set
the renormalization and factorization scales equal to each other
and consider three values $\mu = [m_t/2,m_t,2 m_t]$.
We obtain
\ba
&& \sigma_{\rm LO} (p \bar p \to (t \to l^+ \nu b)\;
+(\bar t \to j_1 j_2 \bar b) + j ) = 36.9^{+22.1}_{-12.8}~{\rm fb};
\nonumber \\
&& \sigma_{\rm NLO} (p \bar p \to (t \to l^+ \nu b)\;
+(\bar t \to j_1 j_2 \bar b) + j ) = 33.6^{-4.0}_{-3.6} ~{\rm fb},
\ea
where the central value refers to $\mu = m_t$, the upper value to
$\mu = m_t/2$ and the lower value to $\mu = 2 m_t$. The improvement
in the scale stability at next-to-leading order
 is apparent.

Our results for the kinematic distributions of the
transverse momentum and the rapidity of the charged lepton $l^+$,
the total transverse energy
$H_\perp$, the transverse momentum and  the rapidity
of the fifth hardest jet and the missing transverse energy
are shown in Fig.~\ref{fig5}(a)-(f), respectively. The bands correspond to choices
of the renormalization and factorization
scales $\mu=[m_t/2,m_t,2m_t]$.  We observe that
the scale variation in rapidity
distributions is in line with the total
cross-section modification and there is little change in shape. However,
for the transverse momenta distributions and the $H_\perp$ distribution,
the situation is more subtle. There are clear differences in
shapes between distributions computed at leading and next-to-leading
order and, in addition,  there are kinematic regions where
the scale variation bands at LO and NLO do not overlap\footnote{This statement
is, of course, contingent upon the chosen range of scales.}.
In particular,
this happens at the high end of the lepton transverse momentum distribution
where, in addition, the scale variation
at NLO actually exceeds the scale variation at leading order.
Similar phenomenon occurs in the missing transverse energy distribution.
On the contrary, the scale variation of
the transverse momentum distribution
of the fifth hardest jet is modest at high $p_{\perp,j}$.

It is interesting that some of the
distributions exhibit peculiar shape distortions. We stress
that these distortions
depend in a very significant way on the choice of
the renormalization and factorizations scales
in leading  order computations and that it is entirely
possible that some of the changes that
we observe in Fig.~\ref{fig5} can  be accommodated by a
kinematic-dependent choice of the renormalization scale.
We will not pursue this topic further in the current paper;
for a related discussion in the context of $W+{\rm jets}$
production, see Refs.~\cite{Bauer:2009km,Berger:2009ep,mz,nlowork}.
For our choices of scales,  we observe
that  the charged lepton transverse momentum  distribution and the
missing transverse energy distribution
become  softer at NLO  when compared to
corresponding leading order distributions.
We can understand this by observing   that
for $\mu = m_t$,   the transverse
momentum distribution of the top quarks becomes softer at NLO as well
\cite{Dittmaier:2007wz,Dittmaier:2008uj}.
On the other  hand,   $\mu = m_t$ is almost the  perfect choice of the
renormalization scale for the  transverse
momentum distribution of the fifth
hardest jet; this  is probably just a numerical coincidence.
While $H_\perp$ distribution does not show significant
distortion at high $H_\perp$, the (broad) peak of the $H_\perp$ distribution
shifts to higher $H_\perp$ values and there is a depletion at low
$H_\perp$ values.\\

Finally, it is clear from Fig.~\ref{fig5} that the positron\footnote{We will refer to the positively charged lepton as the ``positron'' in what follows,
but everything that is said applies to positively charged muons as
well.} rapidity
distribution becomes much more symmetric at next-to-leading order.
To quantify this, we
compute the positron forward-backward asymmetry
\be
A_{e^+} = \frac{\sigma(y_{e^+} > 0)
- \sigma(y_{e^+} < 0)}{\sigma(y_{e^+}> 0)
+ \sigma(y_{e^+} < 0)},\;\;\;
y_{e^+} = \frac{1}{2} \ln \left( \frac{E{_{e^+}}+p_{z,e^+}}{E_{e^+}-p_{z,e^+}}\right).
\label{eq3456}
\ee
For the set up described above, we find the  positron asymmetry
to be strongly reduced at next-to-leading order, similar to the
top quark asymmetry \cite{Dittmaier:2007wz,Dittmaier:2008uj}.
We obtain
\be
A_{e^+}^{\rm LO}  =  -5.05~ \%,~~~~~~~
A_{e^+}^{\rm NLO}  =  -0.5~ \%.
\ee

At first sight,  the strong reduction in the
positron forward-backward asymmetry
observed in $p \bar p \to t \bar t j$ process at NLO
is rather puzzling and worrisome since it suggests a breakdown of
the perturbative expansion for  this quantity.
We will now argue that
{\it i}) these worries are unfounded;
{\it ii}) the large reduction in the forward-backward asymmetry
in $p \bar p \to t \bar t j$ at NLO  is {\it natural};  and
{\it iii}) the NLO asymmetry is, most likely, stable against yet higher
order corrections.   To build up the argument,
we consider the limit of the low jet transverse
momentum cut $p_{\perp j} \to 0$.

We notice that, by requiring  additional jet in the final state
and by taking $p_{\perp j}$ to be small,
we  introduce two types of degrees
of freedom: soft degrees of freedom, controlled by the
jet transverse momentum cut $p_{\perp j}$  and  hard degrees of
freedom,  controlled  by the mass of the top quark.
The crux of our argument is that
the asymmetry in $t \bar t j$ can be generated by {\it both}
{\it soft} and {\it hard} degrees  of freedom but those mechanisms
appear at two consecutive orders in perturbation theory.
The ``soft'' asymmetry appears at leading order while the ``hard'' asymmetry
appears at next-to-leading order.
As evident from their very different
dependence on $p_{\perp,j}$, these two mechanisms are unrelated.
Hence, the dramatic change in the asymmetry at next-to-leading
order should not be taken
as an indicator that the perturbative expansion for this
observable breaks down.

We now explain this argument in detail.
We discuss the top quark  asymmetry rather than the positron
asymmetry, for simplicity. In the limit of small $p_{\perp j}$,
the leading order asymmetry is generated when the
 soft gluon between initial ($ q \bar q$) and final ($t \bar t$)
states is exchanged.
Since the interference diagrams are non-singular  in the collinear
limit, the difference of the forward and backward cross-sections at leading
order can only be proportional to a {\it single logarithm} of the infra-red
cut-off
\be
[\sigma(y_{t} > 0)
- \sigma(y_{t} < 0)]_{\rm LO} \approx \sigma_{A}
\ln\left( \frac{m_t}{p_{\perp,j}}\right),\;\;\
y_{t} = \frac{1}{2} \ln\left( \frac{E_{t}+p_{t,z}}{E_{t}-p_{t,z}}\right).
\ee
Here,  $\sigma_A$ is some quantity with the dimension of the cross-section
and $y_t$ is the top quark rapidity.
On the other hand, the total cross-section that appears in the denominator
in the definition
of the asymmetry  Eq.(\ref{eq3456})
depends on the {\it double logarithm}
of the jet transverse momentum cut
\be
\sigma(y_{t} > 0)
+ \sigma(y_{t} < 0) \sim
\frac{ 2 C_\mathrm{F} \alpha_s}{\pi} \ln^2
\left (\frac{m_t}{p_{\perp,j}} \right )\; \sigma_{t \bar t}.
\label{stot}
\ee
In Eq.(\ref{stot}), $\sigma_{t \bar t} $ is the production cross-section for
$p \bar p \to t \bar t$.  As the result
\be
A_{t\bar tj}^{\rm LO}(p_{\perp,j}) = \frac{\sigma(y_{t} > 0)
- \sigma(y_{t} < 0)}{\sigma(y_{t} > 0)
+ \sigma(y_{t} < 0)} \sim  \ln^{-1} \left(
 \frac{m_t}{p_{\perp,j}} \right ).
\label{lo_ass}
\ee
Hence, we see that at leading order in perturbative QCD,
the asymmetry in $p \bar p \to t \bar t j$
is generated by   soft degrees of freedom and that this asymmetry
{\it decreases }
with the decrease in the jet transverse momentum cut $p_{\perp, j}$.

\begin{figure}[!t]
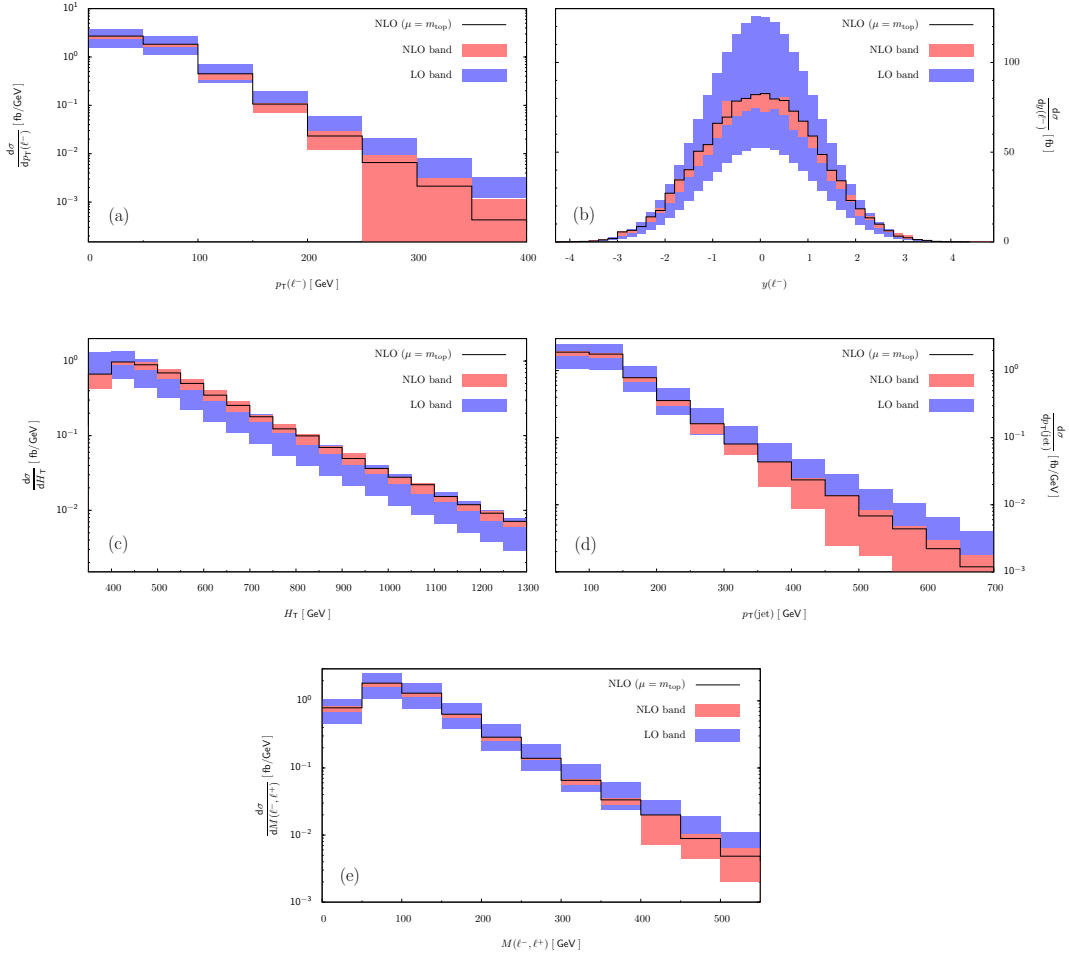

 \begin{center}
  \scalebox{0.4}{\input{LHC13_Fig01.tex}}
  \scalebox{0.4}{\input{LHC13_Fig02.tex}} \\[8mm]
  \scalebox{0.4}{\input{LHC13_Fig04.tex}}
  \scalebox{0.4}{\input{LHC13_Fig06.tex}} \\[8mm]
  \scalebox{0.4}{\input{LHC13_Fig05.tex}}
\end{center}
\caption{
Various distributions for the process
$p p \to (t\rightarrow l^+ \nu b) + (\bar t\rightarrow l^-\bar \nu \bar b)+ j$
at the LHC ($\sqrt{s}=7~{\rm TeV}$)
at leading (blue)
and next-to-leading order (red) in perturbative QCD.
The bands correspond to the choice of the renormalization
and factorization scales $\mu = [m_t/2, m_t, 2m_t]$.
We show distributions of the transverse momentum (a) and  rapidity (b)
of the  positron, the total transverse energy $H_\perp$ (c),
the transverse momentum distribution  of the third hardest jet (d)
and the invariant  mass of the two leptons (e).
}
\label{fig6}
\end{figure}

At next-to-leading order, the asymmetry can still be generated by soft
exchanges. In such a case, it is natural to expect that {\it moderate}
correction to the leading order asymmetry Eq.(\ref{lo_ass})
is generated.  However, it is interesting  that at
next-to-leading order,
a new mechanism for generating the asymmetry
in $p \bar p \to t \bar t j $  appears. Indeed,
we can use  hard degrees of freedom to generate the
asymmetry, in much the same way as it is generated in the inclusive
$p \bar p \to t \bar t $ process.  In addition, we can
 use the emission
of an additional  soft and collinear  jet from the initial
state to provide
conventional {\it double logarithmic} enhancement of this ``hard''
asymmetry.
This mechanism leads to the following forward-backward
cross-section difference in Eq.(\ref{eq3456})
as
\be
\sigma(y_{t} > 0)
- \sigma(y_{t} < 0)
\sim \frac{2 C_\mathrm{F} \alpha_s}{\pi}
\ln^2 \left ( \frac{m_t}{p_{\perp,j}} \right )\;
A_{t\bar t} \; \sigma_{t \bar t},
\label{eq49}
\ee
where $A_{t \bar t}$ is the $p \bar p \to t \bar t $
{\it inclusive} asymmetry. Taking the ratio of
Eq.(\ref{eq49}) and Eq.(\ref{stot}), we find
\be
A_{t\bar tj}^{\rm hard} \approx A_{t \bar t}.
\ee

The full forward-backward asymmetry at next-to-leading order is given
by the sum of the two mechanisms
\be
A_{t\bar tj}^{\rm NLO} \approx A_{t\bar tj}^{\rm soft}(p_{\perp,j})
+ A_{ t \bar t j}^{\rm hard}.
\label{eq1123}
\ee
Following the preceding discussion, we estimate
\be
A_{t\bar tj}^{\rm soft}(p_{\perp,j}) \sim A_{t\bar tj}^{\rm LO}(p_{\perp,j})
\sim \ln^{-1} \left ( \frac{m_t}{p_{\perp,j}} \right ),\;\;\;
A_{t\bar tj}^{\rm hard} \approx A_{t \bar t}.
\label{eq0123}
\ee
Numerically,  the ``soft'' asymmetry  at $p_{\perp,j}=30~{\rm GeV}$
is $A_{t\bar tj}^{\rm LO}(30~{\rm GeV}) \sim -7 \%$ and  the ``hard'' asymmetry
is  $A_{t \bar t} \sim 5 \%$ \cite{kuhn},  so
that the significant reduction in the leading order
$p \bar p \to t \bar t j$ asymmetry is observed,
once the NLO QCD corrections are accounted for.
A  peculiar consequence of this argument
is that, in the limit of a very small jet transverse momentum
cut, the top quark forward-backward asymmetries
in the $p \bar p \to t \bar  t j$ process and in the
inclusive $p \bar p \to t \bar t$  process coincide
\be
\lim_{p_{\perp,j} \to 0}  A_{t\bar tj}(p_{\perp,j}) \to A_{t \bar t},
\ee
because  the ``soft'' component of the asymmetry vanishes as the
inverse logarithm of the cut on the jet  transverse momentum.

We stress that the  observation by Dittmaier, Uwer and Weinzierl
of the  large NLO QCD corrections
to the top quark forward-backward asymmetry in $p \bar p \to t \bar t j$
\cite{Dittmaier:2007wz,Dittmaier:2008uj}
is important outside the context of that computation. Indeed, since
the asymmetry in $p\bar p \to t \bar t$ appears at NLO and since NLO
is the highest order in the perturbative expansion available for that
process, corrections to the inclusive asymmetry are {\it not known}.
On the other hand, because the asymmetry in $ p\bar p  \to
t \bar t j$ appears {\it already}
at leading order, a NLO QCD computation for
$p\bar p  \to t \bar t j$
gives {\it corrections} to the leading order forward-backward
asymmetry. The fact that
these  corrections turn out to be nearly $100 \%$,
as discovered in Refs.~\cite{Dittmaier:2007wz,Dittmaier:2008uj},
leads to doubts about the robustness of the existing predictions
for the inclusive asymmetry in the $p \bar p \to t \bar t$ process.
Of course,  the issue of robustness is rather important
given the existing discrepancy between
the measurement of the asymmetry in $p \bar p \to t \bar t$
\cite{ass_cdf,ass_d0} and
the theoretical prediction  \cite{kuhn}.

We believe that our argument supports the robustness
of the theoretical prediction of the
inclusive asymmetry $A_{t \bar t}$. In fact, as we have
explained, there are two physically distinct mechanisms that generate  the
asymmetry in $p \bar p \to t \bar t j$. These mechanisms are related
to the existence of two types of degrees of freedom,
which can be clearly separated by lowering the
cut on the jet transverse momentum.
The only peculiarity  about the asymmetry is that the
 two mechanisms for the asymmetry generation
do not appear at the same order of the perturbative expansion which
leads to apparent problems with the interpretation of its convergence.
Needless to say that we do not see any other mechanism that can start
contributing to the asymmetry  at next-to-next-to-leading order (NNLO)
and beyond. Therefore,
we believe,  that the NLO prediction
for the asymmetry in $p \bar p \to t \bar t j$ is robust.  Similarly,
since inclusive $p \bar p \to t \bar t$ production process
is only sensitive to hard degrees of freedom, no new mechanism
for the asymmetry generation  can appear at
NNLO and beyond. Therefore, we
believe that the existing prediction for the top quark forward-backward asymmetry
in $p \bar p \to t \bar t$ is robust as well.

\begin{figure}[!t]
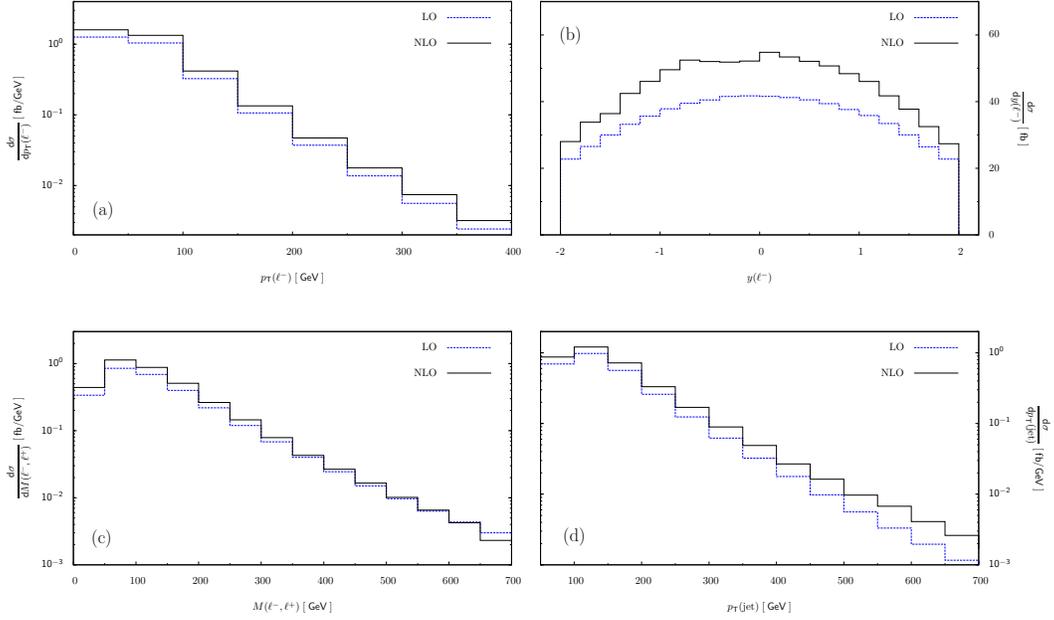

 \begin{center}
  \scalebox{0.4}{\input{LHC14x_Fig01.tex}}
  \scalebox{0.4}{\input{LHC14x_Fig02.tex}} \\[8mm]
  \scalebox{0.4}{\input{LHC14x_Fig04.tex}}
  \scalebox{0.4}{\input{LHC14x_Fig05.tex}}
\end{center}
\caption{
Kinematic distributions for the process
$p p \to (t\to l^+ \nu b) + (\bar t\to l^-\bar \nu \bar b)+ j$
at the LHC ($\sqrt{s} = 14~{\rm TeV}$) with
weak boson fusion cuts. The renormalization and
factorization scales are set to $\mu = m_t$. Distributions at
leading (blue)
and next-to-leading (black) order in perturbative QCD are shown.
We show distributions of the transverse momentum  (a) and
rapidity of the electron  (b), the invariant mass of the electron
and the positron (c), the transverse momentum of the leading jet (d).
}
\label{fig7}
\end{figure}

\subsection{ $t \bar t + {\rm jet}$ production in dilepton channel at the LHC }

We now turn our attention to the LHC, $\sqrt{s} = 7~{\rm TeV}$. We
consider  dilepton channel and require
three or more jets with
$p_{\perp,j} > 50~{\rm GeV}$, $p_{\perp,\mathrm{b-jet}} > 20~{\rm GeV}$.
Our definition of the $b$-jet is explained above.
We also apply  the following cuts  on the
lepton transverse momentum
$p_{\perp,l} > 20~{\rm GeV}$ and
the missing transverse momentum $p_{\perp,\rm miss} > 40~{\rm GeV}$.
We begin by showing
the cross-sections for the three values of the factorization
and renormalization scales $\mu = [m_t/2, m_t, 2m_t]$.
We find
\ba
&& \sigma_{\rm LO} (p p \to (t \to l^+ \nu b)\;
+(\bar t \to l^- \bar \nu \bar b) + j ) = 229.9^{+133.7}_{-78.2}~{\rm fb};
\nonumber \\
&& \sigma_{\rm NLO} (p p \to (t \to l^+ \nu b)\;
+(\bar t \to l^- \bar \nu \bar b) + j ) = 256.5^{-14.8}_{-25.6}~{\rm fb},
\label{eq_lhc}
\ea
where the central value refers to $\mu = m_t$, the upper value to
$\mu = m_t/2$ and the lower value to $\mu = 2 m_t$.  Similar to
the Tevatron case, the NLO cross-section is very stable against changes
in the renormalization and factorization scales, in contrast to the leading
order result.

Our results for kinematic distributions are displayed  in
Fig.~\ref{fig6}, where we present  the transverse momentum and  the rapidity
distributions of the positively charged lepton, the distribution
of the total transverse energy $H_\perp$,
the  transverse momentum distribution  of the third hardest jet
and the distribution of the invariant  mass of the two leptons.
We note that for some choices of the renormalization and
factorization scales, the NLO results for the lepton
transverse momentum distribution become negative at
high $p_\perp(l^+) \gsim  400~{\rm GeV}$.
 This indicates that
$\mu = m_t/2$ is too low a scale for this observable and that
there is large residual scale dependence in this observable even
at next-to-leading order.
The $H_\perp$-distribution
is shifted to higher values, leading to
positive corrections in the high-$H_\perp$ tail. The transverse momentum
distribution of the third jet becomes softer
and the lepton invariant mass distribution is only marginally
affected. Many of these features can probably be understood
by performing a leading order computation with the kinematic-dependent
choices of the scales. As we already mentioned in the context of the
Tevatron discussion, we do not pursue  this topic in the
present paper and reserve it for future work.\\

\begin{figure}[!t]
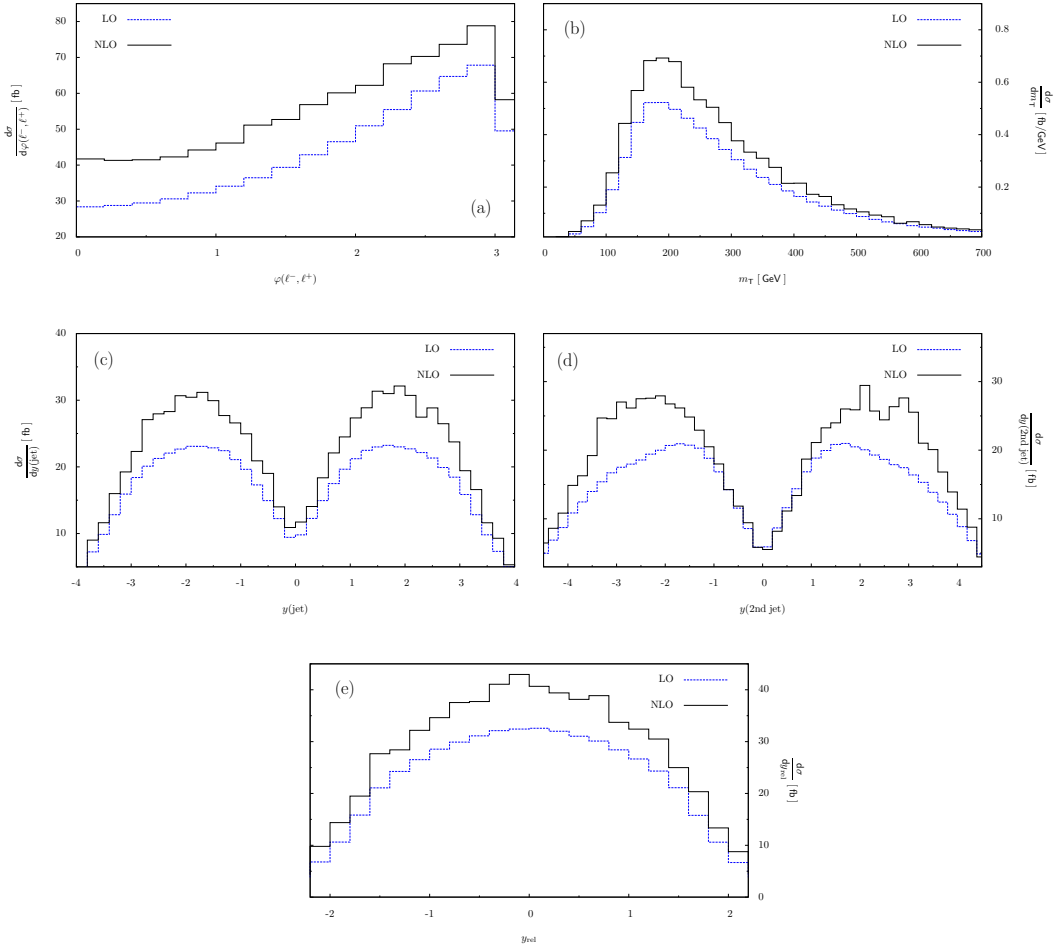

 \begin{center}
  \scalebox{0.4}{\input{LHC14x_Fig06.tex}}
  \scalebox{0.4}{\input{LHC14x_Fig07.tex}} \\[8mm]
  \scalebox{0.4}{\input{LHC14x_Fig08.tex}}
  \scalebox{0.4}{\input{LHC14x_Fig09.tex}} \\[8mm]
  \scalebox{0.4}{\input{LHC14x_Fig10.tex}}
\end{center}
\caption{
Kinematic distributions for the process
$p p \to (t\to l^+ \nu b) + (\bar t \to l^-\bar \nu \bar b)+ j$
at the LHC ($\sqrt{s} = 14~{\rm TeV}$) with
weak boson fusion cuts. The renormalization and
factorization scales are set to $\mu = m_t$. Distributions at
leading (blue)
and next-to-leading (black) order in perturbative QCD are shown.
We show distributions relevant for the discrimination between
the signal and the background. In particular, we display distributions
of  the relative azimuthal angle of the electron and the positron (a),
the $m_{T}$ variable (b) (see text),
the rapidity of the leading jet (c), the rapidity of the
second leading jet (d)
 and the relative rapidity
of the ``veto'' (third hardest)  jet (e). For the
definition of the relative rapidity of the veto jet, see text.
}
\label{fig8}
\end{figure}

Our last example concerns NLO QCD corrections to $pp \to t \bar tj$
process
in the kinematic region relevant for the Higgs boson searches in
weak boson fusion  $pp \to jjH \to jjW^+W^-$
production channel, with leptonic decays
of the $W$-bosons. It was shown in Ref.~\cite{eboli} that the
$pp \to t \bar t j$ process is the largest background to the
weak boson fusion signal. For our  calculation, we  take the
LHC energy to be $\sqrt{s} = 14~{\rm TeV}$.
The event selection is based on the
following cuts \cite{eboli}.
We define jets with the usual $k_\perp$ algorithm,
$R_{ij} = 0.5$ and the transverse momentum cut
$p_{\perp,j} > 20~{\rm GeV}$.
The two jets with the highest transverse
momentum (tag jets)  are required to
satisfy $p_\perp^{(1)} > 40~{\rm GeV}$ and
$p_\perp^{(2)} > 20~{\rm GeV}$. Those jets should be in the opposite
hemispheres $y_1 y _2 < 0$ and widely separated in rapidity
$|y_1 - y_2| > 3.0$. The invariant mass of the two tag jets
should be large $m_{j_1 j_2} > 550~{\rm GeV}$.
The two leptons are required to have large
transverse momenta $p_{\perp,l} > 20~{\rm GeV}$ and central
rapidities  $|y_l| < 2$. In principle, there are other cuts
that are imposed to separate  the weak boson fusion signal
but we do not employ those  cuts in what follows.  For the cross-sections
at $\mu = m_t$  we find
\ba
&& \sigma_{\rm LO} (p  p \to (t \to l^+ \nu b)\;
+(\bar t \to l^- \bar \nu \bar b) + j ) = 139.6~{\rm fb};
\nonumber \\
&& \sigma_{\rm NLO} (p  p \to (t \to l^+ \nu b)\;
+( \bar t \to l^- \bar \nu \bar b) + j ) = 177.9~{\rm fb}.
\label{tc}
\ea
The next-to-leading order cross-section exceeds the leading order
cross-section by twenty-five percent which is not  unusual.
In Fig.~\ref{fig7}
we show a number of kinematic distributions for the weak boson
fusion cuts.  In particular, lepton transverse momentum and rapidity
distributions, distribution of the dilepton invariant mass and the transverse
momentum distribution of the hardest jet are displayed.
In Fig.~\ref{fig8},  we show  kinematic distributions  that
can be used to discriminate between the $t \bar t + {\rm jet}$
background and the weak boson fusion signal. We display the
distribution in the relative azimuthal angle of the two
leptons,   the distribution of the approximate transverse
mass of the two $W$-bosons $m_{T}$ and some rapidity distributions.
The definition of the $m_{T}$-variable is given in Ref.~\cite{eboli}
\be
m_{T} =
\left [ \left (E_{\perp,l^+l^-} + {\tilde E}_{\perp,{\rm miss}} \right )^2
- \left (\vec p_{\perp,l^+l^-} + \vec p_{\perp,{\rm miss}} \right )^2
\right ]^{1/2}.
\ee
In this formula,
$p_{\perp,{\rm miss}}$ is the missing transverse momentum which
we associate with the vector sum of the momenta of the two neutrinos
and ${\tilde E}_{\perp,\rm miss} = \sqrt{p_{\perp,{\rm miss}}^2 + m_{l^+l^-}^2}$.
As explained in Ref.~\cite{eboli}, $m_{T}$ is a good  approximation
to the actual transverse mass of the Higgs boson, provided that
the Higgs mass is betwen $160$ and $200$~GeV.
A glance at
Figs.~\ref{fig7},~\ref{fig8} suggests that shapes of lepton $p_\perp$ and
rapidity distributions and the dilepton invariant mass distribution are
not strongly affected by the radiative corrections for $\mu = m_t$.
The distribution of the transverse momentum
of the leading jet becomes harder.
The shape of $m_T$ distribution is not changed by the NLO QCD
corrections but there appears  to be a
shape change in the distribution of the relative azimuthal angle
of the two leptons. As the result,  the $\Delta \phi$ distribution
at next-to-leading order is less peaked at $\Delta \phi \approx \pi$,
compared to
the leading order result.  We also plot two rapidity distributions.
The rapidity distribution of the hardest jet at NLO (Fig.~\ref{fig8}$(c)$)
is well reproduced
by the re-scaling of the leading order distribution.  On the other
hand, the NLO rapidity
distribution of the second hardest jet, shown
in Fig.~\ref{fig8}$(d)$,
is shifted to larger absolute
values and the shape of the distribution changes.
The distribution of the relative rapidity
$ y _{\rm rel} = y_{\rm veto} - 0.5 \left (y_{j_1} + y_{j_2} \right )$
of the ``veto jet'' is also shown,  Fig.~\ref{fig8}$(e)$. The veto jet is
 defined as the jet between the two tagged jets. The shape of
$y_{\rm rel}$ distribution does not change from leading to
next-to-leading order.

\section{Conclusions}
\label{sectc}
We describe the calculation of the NLO QCD corrections to the
production of a $t \bar t$ pair in association with
a hard jet at the Tevatron and the LHC.   The one-loop virtual
corrections are computed using the framework of generalized
$D$-dimensional unitarity. We present numerical
results for one-loop helicity amplitudes for the processes $0 \to t \bar t ggg$ and
$0 \to t \bar t q \bar q g$.

Availability of helicity amplitudes enables us to include
decays of top quarks, in the narrow width
approximation,  at almost  no additional cost. We
do that in leading order in perturbative QCD. We
account for all the spin correlations
between top quark decay products exactly and
include constraints on top quark
decay products in the computation.
We emphasize that these features may be important
for the detailed comparison between the theoretical predictions for
$pp(p \bar p) \to t \bar t + {\rm jet}$ and results of experimental
measurements as well as for the the understanding of the $t \bar t + {\rm jet}$
process in the kinematics relevant for New Physics searches.

We consider production of the $t \bar t$ pair in
association with one hard jet
both at the Tevatron and the LHC,
for realistic cuts on the final state leptons, jets and missing energy.
For the LHC, we also investigate NLO QCD corrections
to  $t \bar t + {\rm jet}$
process in the kinematic region selected by the weak boson fusion
cuts \cite{eboli}.
For all the cases considered,  we find  reduced dependence
on the renormalization and  factorization scales. This is true
for total cross-sections and main regions of kinematic distributions.
For tails of kinematic distributions, the situation is more subtle
since next-to-leading order and leading order scale-dependence bands
are often similar and may  not even overlap.

We confirm findings of Refs.~\cite{Dittmaier:2007wz, Dittmaier:2008uj}
that the top quark forward-backward asymmetry in the $p \bar p \to t \bar t j$ process
receives very large corrections at next-to-leading order. We explain
the origin of these large corrections
and argue  that they should have been expected.
Since there is no mechanism, that would lead to large corrections
to top quark forward-backward asymmetry in $p\bar p \to t \bar t j$
process  in yet higher orders, we consider
the NLO QCD prediction for the asymmetry as quite reliable. We note that
large NLO QCD corrections
to the asymmetry in $p\bar p \to t \bar t j$ process are  particular to the
presence of a jet in the final state; the enhancement  mechanism does not
work  for the inclusive asymmetry in
$p\bar p \to t \bar t$. Therefore, we do not expect that missing
higher order corrections are  the reason for the significant discrepancy
between the QCD prediction for the forward-backward asymmetry
\cite{kuhn} and the experimental results \cite{ass_cdf,ass_d0}.

The calculation reported in this paper opens up a possibility
to study production of a $t \bar t$ pair in association with
a hard jet
at hadron colliders in a realistic manner.
Further improvements may
require the inclusion of the QCD corrections to top quark decays.
In addition, there is significant  interest in other
processes where a $t \bar t$ pair is produced in an association
with either a vector boson or the Higgs boson \cite{ttg, ttz,tth}.
It will be definitely interesting to refine NLO QCD predictions
for those processes to account for the decays of top quarks.

{\bf Acknowledgments}
We are grateful  to Zoltan Kunszt for his comments on the manuscript.
This research  is supported by the NSF under grant
PHY-0855365 and by the startup
funds provided by Johns Hopkins University.
Calculations reported in this paper were performed on the Homewood
High Performance Cluster of Johns Hopkins University.

\newpage

\appendix

\section{}
In this Appendix, we give explicit  expressions \cite{egkmz} for
color-ordered amplitudes
$B^{[1]}_{i}$ and $B^{[1/2,q(t)]}_{i}$ in terms
of the primitive amplitudes. The different classes of the amplitudes
are shown in Fig.~\ref{fig3}.
For the amplitudes with  closed
fermion loop we find
\begin{eqnarray}
&& B^{[1/2,q(t)]}_{1} = -A^{[1/2,q(t)]}_L(1_\tb, 5_g,4_q,3_\qb,2_t),\;\;\;
 B^{[1/2,q(t)]}_{2} = -A^{[1/2,q(t)]}_L(1_\tb, 4_q,3_\qb,2_t,5_g),
\nn \\
&& B^{[1/2,q(t)]}_{3} = -A^{[1/2,q(t)]}_L(1_\tb, 4_q,3_\qb,5_g,2_t),\;\;\;
B^{[1/2,q(t)]}_{4} = -A^{[1/2,q(t)]}_L(1_\tb, 4_q,5_g,3_\qb,2_t).
\end{eqnarray}

There are three classes of primitive amplitudes that we need to consider
for four-quark processes without closed fermion loops
\be
B^{[1]}_{i} = B^{[1],a}_{i}+
B^{[1],b}_{i} + B^{[1],c}_{i}\,,
\ee
they are shown in Fig.~\ref{fig3}(a)-(c).
Amplitudes from each class are written as linear combinations of
primitives amplitudes.

For the class ``a'', we find
\begin{eqnarray}
B^{[1],a}_{1} &=&
\left ( 1- \frac{1}{N_c^2} \right )
A_L^{[1],a}\left (1_\tb, 2_t, 3_\qb, 4_q, 5_g \right )
- \frac{1}{N_c^2}
\left (
-A_L^{[1],a}\left (1_\tb, 5_g, 2_t, 3_\qb, 4_q \right )
\right.  \\
&& \left .
-A_L^{[1],a} \left (1_\tb, 5_g,2_t,  4_q, 3_\qb  \right )
-A_L^{[1],a} \left (1_\tb,  2_t,5_g, 3_\qb, 4_q \right )
-A_L^{[1],a} \left (1_\tb, 2_t, 5_g, 4_q,3_\qb  \right )
\right. \nn \\
&& \left .
-A_L^{[1],a} \left (1_\tb, 2_t, 3_\qb,5_g,  4_q  \right )
+A_L^{[1],a} \left (1_\tb, 2_t,  4_q, 5_g, 3_\qb \right )
+A_L^{[1],a} \left (1_\tb, 2_t, 4_q, 3_\qb , 5_g \right )
\right ), \nn \\
B^{[1],a}_{2} &=&
 A_L^{[1],a} \left (1_\tb, 2_t, 5_g, 4_q,3_\qb  \right )
+ A_L^{[1],a} \left (1_\tb, 2_t,  4_q,5_g,3_\qb  \right )
+ A_L^{[1],a} \left (1_\tb, 2_t,  4_q,3_\qb,5_g  \right )
\nn \\
&&
-\frac{1}{N_c^2} \left (
A_L^{[1],a} \left (1_\tb, 5_g, 2_t, 3_\qb,4_q  \right )
+ A_L^{[1],a} \left (1_\tb, 5_g, 2_t, 4_q, 3_\qb  \right )
\right ),
\\
B^{[1],a}_{3} &=&
 \left ( 1- \frac{1}{N_c^2} \right )
A_L^{[1],a}\left (1_\tb, 2_t, 5_g, 3_\qb, 4_q \right )
- \frac{1}{N_c^2}
\left (
-A_L^{[1],a}\left (1_\tb, 5_g, 2_t, 3_\qb, 4_q \right )
\right.  \\
&& \left .
-A_L^{[1],a} \left (1_\tb, 5_g,2_t,  4_q, 3_\qb  \right )
-A_L^{[1],a} \left (1_\tb,  2_t, 3_\qb, 5_g, 4_q \right )
+A_L^{[1],a} \left (1_\tb, 2_t, 4_q,5_g, 3_\qb  \right )
\right. \nn \\
&& \left .
-A_L^{[1],a} \left (1_\tb, 2_t, 3_\qb,  4_q,5_g  \right )
-A_L^{[1],a} \left (1_\tb, 2_t,  4_q, 3_\qb,5_g \right )
+A_L^{[1],a} \left (1_\tb, 2_t,5_g, 4_q, 3_\qb\right )
\right ), \nn \\
B^{[1],a}_{4} &=&
- A_L^{[1],a} \left (1_\tb, 5_g, 2_t, 4_q,3_\qb  \right )
- A_L^{[1],a} \left (1_\tb, 2_t, 5_g,  4_q,3_\qb  \right )
- A_L^{[1],a} \left (1_\tb, 2_t,  4_q,3_\qb,5_g  \right )
\nn \\
&&
-\frac{1}{N_c^2} \left (
 A_L^{[1],a} \left (1_\tb, 2_t,3_\qb,5_g, 4_q  \right )
-A_L^{[1],a} \left (1_\tb,  2_t,4_q, 5_g, 3_\qb  \right )
\right ).
\end{eqnarray}

For classes ``b'' and ``c'' we obtain
\begin{eqnarray}
B^{[1],b}_{1} &=&
\frac{1}{N_c^2}
A_L^{[1],b} \left (1_\tb, 5_g, 4_q,3_\qb, 2_t  \right ),
\nn \\
B^{[1],b}_{2} &=&
-\frac{1}{N_c^2}
\left (
A_L^{[1],b} \left (1_\tb, 5_g, 4_q,3_\qb, 2_t  \right )
+A_L^{[1],b} \left (1_\tb, 4_q,5_g,3_\qb, 2_t  \right )
+A_L^{[1],b} \left (1_\tb, 4_q, 3_\qb,5_g,2_t  \right )
\right ),
\nn \\
B^{[1],b}_{3} &=&
\frac{1}{N_c^2}
A_L^{[1],b} \left (1_\tb, 4_q,3_\qb,5_g, 2_t  \right ),
\nn \\
B^{[1],b}_{4} &=&
-A_L^{[1],b} \left (1_\tb, 5_g, 4_q,3_\qb, 2_t  \right )
-A_L^{[1],b} \left (1_\tb, 4_q,3_\qb,5_g, 2_t  \right )
-A_L^{[1],b} \left (1_\tb, 4_q,3_\qb, 2_t,5_g  \right )
\nn \\
&&
-\left ( 1 - \frac{1}{N_c^2}
\right )
A_L^{[1],b} \left (1_\tb, 4_q,5_g,3_\qb, 2_t  \right ),
\end{eqnarray}

\begin{eqnarray}
B^{[1],c}_{1} &=&
\frac{1}{N_c^2}
A_L^{[1],c} \left (1_\tb, 5_g, 4_q,3_\qb, 2_t  \right ),
\nn \\
B^{[1],c}_{2} &=&
-A_L^{[1],c} \left (1_\tb,5_g, 4_q, 3_\qb,2_t \right )
-A_L^{[1],c} \left (1_\tb, 4_q,5_g,3_\qb, 2_t  \right )
-A_L^{[1],c} \left (1_\tb, 4_q,3_\qb, 5_g,2_t  \right )
\nn \\
&&
-\left (1-\frac{1}{N_c^2} \right )
A_L^{[1],c} \left (1_\tb, 4_q,3_\qb, 2_t,5_g  \right ),
\nn \\
B^{[1],c}_{3} &=&
\frac{1}{N_c^2}
A_L^{[1],c} \left (1_\tb, 4_q,3_\qb,5_g, 2_t  \right ),
\\
B^{[1],c}_{4} &=&
-\frac{1}{N_c^2} \left (
A_L^{[1],c} \left (1_\tb, 5_g, 4_q,3_\qb, 2_t  \right )
+A_L^{[1],c} \left (1_\tb, 4_q,3_\qb,5_g, 2_t  \right )
+A_L^{[1],c} \left (1_\tb, 4_q,3_\qb, 2_t,5_g  \right )
\right ). \nn
\end{eqnarray}

\section{}

In the Appendix we give sample results for the
helicity amplitudes for the kinematic
point considered in Ref.~\cite{Dittmaier:2008uj}. The momenta for the reaction
$ab \to t \bar t + c$ read (in units of GeV)
\begin{small}
\ba
&&   p_a = (500,0,0,500), \nn \\
&&  p_b = (500,0,0,-500), \nn \\
&& p_t = (458.53317553852783,207.0255169909440,0,370.2932732896167),\nn \\
&& p_{\bar t} = (206.6000026080000,-10.65693677252589,
42.52372780926147,-102.39982104210421085), \nn \\
&& p_c = (334.8668220067217,-196.3685802184181,-42.52372780926147,-267.8934522475083). \nn
\ea
\end{small}
We present results for the ratio of one-loop helicity
left-primitive amplitudes to the
corresponding tree primitive amplitudes in Tables B1-B4.
Those results do not include external wave function renormalization constants and the coupling constant renormalization.
However, mass counterterms are included to obtain a gauge-invariant result.
We define
\be
r^{[j]}(i_1,i_2,i_3,i_4,i_5) = \frac{1}{c_\Gamma} \frac{A_L^{[j]}(i_1,i_2,i_3,i_4,i_5)}{A^{\rm tree}(i_1,i_2,i_3,i_4,i_5)}.
\ee

The one-loop scattering amplitudes presented below are computed for
definite helicity states of external quarks and gluons.
For a gluon with momentum $p = E(1,\sin \theta \cos \phi,
\sin \sin \phi, \cos \theta)$, we define polarization vectors
as
\be
\epsilon_\mu^{\pm}(p) = \frac{1}{\sqrt{2}} \left (
0,\cos \theta \cos \phi \mp i \sin \phi,\cos \theta \sin \phi \pm i \cos \phi,-\sin \phi \right ).
\ee
We use the Dirac representation of the $\gamma$-matrices. For a fermion
with the on-shell momentum $p = (E,p_x,p_y,p_z)$, $p^2 = m^2$, and
helicity $\lambda = \pm$,  we
use the spinors
\ba
&& u_+(p) =
\left (
\begin{array}{c}
\sqrt{E+m}  \\
0 \\
p_z/\sqrt{E+m} \\
(p_x+ip_y)/\sqrt{E+m}
\end{array}
\right ),
\;\;\;\;
u_-(p) =
\left (
\begin{array}{c}
0 \\
\sqrt{E+m}  \\
(p_x-ip_y)/\sqrt{E+m}
\\
-p_z/\sqrt{E+m}
\end{array}
\right ),  \\
&& v_+(p) =
\left (
\begin{array}{c}
p_z/\sqrt{E+m}
\\
(p_x+ip_y)/\sqrt{E+m}
\\
\sqrt{E+m}  \\
0
\end{array}
\right ),
\;\;\;\;
v_-(p) =
\left (
\begin{array}{c}
(p_x-ip_y)/\sqrt{E+m}
\\
-p_z/\sqrt{E+m}
\\
0 \\
\sqrt{E+m}
\end{array}
\right ).
\ea
The case of massless leptons is obtained by taking $m \to 0$ limit
in the above formulas.

\newpage

\begin{table}[h]
\begin{center}
\begin{tabular}{| c|c|c|c| }
\hline
Helicity amplitude&$\;1/\epsilon^2\;$&$1/\epsilon$&$\epsilon^0$\\ \hline
$A^{\rm tree}(1_{\bar{t}}^{+},2_{t}^{+},3_{g}^{-},4_{g}^{-},5_{g}^{+})$ & & &   $-0.11654501+   0.74382743\mathrm{i} $\\
$r^{[1]}(1_{\bar{t}}^{+},2_{t}^{+},3_{g}^{-},4_{g}^{-},5_{g}^{+})$ & $-3.000000$ & $   6.86872922   -6.28318531 \mathrm{i} $ & $   8.70123544+   3.76992817 \mathrm{i}$ \\ \hline
$A^{\rm tree}(1_{\bar{t}}^{+},3_{g}^{-},2_{t}^{+},4_{g}^{-},5_{g}^{+})$  & & & $   0.13444688   -0.82987444\mathrm{i} $ \\
$r^{[1]}(1_{\bar{t}}^{+},3_{g}^{-},2_{t}^{+},4_{g}^{-},5_{g}^{+})$ & $-2.000000$ & $   5.61128078   -3.14159266 \mathrm{i} $ & $   7.23821384   -1.35596662 \mathrm{i}$ \\ \hline
$A^{\rm tree}(1_{\bar{t}}^{+},3_{g}^{-},4_{g}^{-},2_{t}^{+},5_{g}^{+})$ & & &  $ -0.01014177+   0.07636266\mathrm{i}$ \\
$r^{[1]}(1_{\bar{t}}^{+},3_{g}^{-},4_{g}^{-},2_{t}^{+},5_{g}^{+})$ & $-1.000000$ & $   4.47200426   -6.28318345 \mathrm{i} $ & $   6.07761708+   8.88451524 \mathrm{i}$ \\ \hline
$A^{\rm tree}(1_{\bar{t}}^{+},3_{g}^{-},4_{g}^{-},5_{g}^{+},2_{t}^{+})$ & & &  $ -0.00776010+   0.00968436\mathrm{i} $ \\
$r^{[1]}(1_{\bar{t}}^{+},3_{g}^{-},4_{g}^{-},5_{g}^{+},2_{t}^{+})$ & $ 0.000000$ & $   2.73111593   -3.22388847 \mathrm{i} $ & $  62.01806062+ 113.31811124 \mathrm{i}$ \\ \hline
\end{tabular}
\end{center}
\caption{Examples of $A_L^{[1]}$ amplitudes contributing to the process $g_3 g_4 \to  t \bar t g_5 $.
The tree level amplitudes $A^{\rm tree}$ are given in units of $(100\, \mathrm{GeV})^{-1}$.
}
\label{table:nterm3}
\end{table}
\begin{table}[h]
\begin{center}
\begin{tabular}{| c|c|c|c| }
\hline
Helicity amplitude&$\;1/\epsilon^2\;$&$1/\epsilon$&$\epsilon^0$\\ \hline

$A^{\rm tree}(1_{\bar{t}}^{+},2_{t}^{+},3_{g}^{-},4_{g}^{-},5_{g}^{+})$ & & &   $-0.11654501+   0.74382743\mathrm{i}$\\
$r^{[1/2,q]}(1_{\bar{t}}^{+},2_{t}^{+},3_{g}^{-},4_{g}^{-},5_{g}^{+})$ & $ 0.000000$ & $   0.00000000   $ & $  -0.06796975   -0.26181130 \mathrm{i}$ \\ \hline
$A^{\rm tree}(1_{\bar{t}}^{+},5_{g}^{+},2_{t}^{+},3_{g}^{-},4_{g}^{-})$  & & & $      0.00077387   -0.12309159\mathrm{i} $ \\
$r^{[1/2,q]}(1_{\bar{t}}^{+},5_{g}^{+},2_{t}^{+},3_{g}^{-},4_{g}^{-})$ & $ 0.000000$ & $   0.00000000 $ & $  -0.11916673   -0.05644255 \mathrm{i}$ \\ \hline
$A^{\rm tree}(1_{\bar{t}}^{+},2_{t}^{+},3_{g}^{-},4_{g}^{-},5_{g}^{+})$ & & &   $-0.11654501+   0.74382743\mathrm{i}$\\
$r^{[1/2,t]}(1_{\bar{t}}^{+},2_{t}^{+},3_{g}^{-},4_{g}^{-},5_{g}^{+})$ & $ 0.000000$ & $   0.00000000  $ & $  -0.10289728   -0.06888369 \mathrm{i}$ \\ \hline
$A^{\rm tree}(1_{\bar{t}}^{+},5_{g}^{+},2_{t}^{+},3_{g}^{-},4_{g}^{-})$  & & & $      0.00077387   -0.12309159\mathrm{i} $ \\
$r^{[1/2,t]}(1_{\bar{t}}^{+},5_{g}^{+},2_{t}^{+},3_{g}^{-},4_{g}^{-})$ & $ 0.000000$ & $   0.00000000 $ & $  -0.13739705+   0.06472332 \mathrm{i}$ \\ \hline
\end{tabular}
\end{center}
\caption{Examples of $A_L^{[1/2]}$ amplitudes  contributing to the process $g_3 g_4 \to  t \bar t g_5 $.
 First and second rows correspond to closed fermion loops with massless fermions; third and fourth rows correspond to closed fermion loops with massive
top quarks.
The tree level amplitudes $A^{\rm tree}$ are given in units of $(100\, \mathrm{GeV})^{-1}$.}
\label{table:nterm3}
\end{table}

\newpage
\begin{table}[h]
\begin{center}
\begin{tabular}{| c|c|c|c| }
\hline
Helicity amplitude&$\;1/\epsilon^2\;$&$1/\epsilon$&$\epsilon^0$\\ \hline
 $A^{\rm tree}(1_{\bar{t}}^{+},2_{t}^{+},3_{\bar{q}}^{-},4_{q}^{+},5_{g}^{-})$ & & & $  -0.03342616   -0.15320512\mathrm{i} $ \\
$r^{[1],a}(1_{\bar{t}}^{+},2_{t}^{+},3_{\bar{q}}^{-},4_{q}^{+},5_{g}^{-})$ & $-2.000000$ & $   5.53799592   -3.14159265 \mathrm{i} $ & $   6.33186822+   0.35033806 \mathrm{i}$ \\ \hline
$A^{\rm tree}(1_{\bar{t}}^{+},5_{g}^{-},2_{t}^{+},3_{\bar{q}}^{-},4_{q}^{+})$  & & & $   0.01451328   -0.07862749\mathrm{i} $ \\
$r^{[1],a}(1_{\bar{t}}^{+},5_{g}^{-},2_{t}^{+},3_{\bar{q}}^{-},4_{q}^{+})$ & $-1.000000$ & $   4.97232917    $ & $   5.61332512   -1.66499812 \mathrm{i}$ \\ \hline
$A^{\rm tree}(1_{\bar{t}}^{+},2_{t}^{+},5_{g}^{-},3_{\bar{q}}^{-},4_{q}^{+})$ & & & $  -0.00409831   -0.01494084\mathrm{i} $ \\
$r^{[1],a}(1_{\bar{t}}^{+},2_{t}^{+},5_{g}^{-},3_{\bar{q}}^{-},4_{q}^{+})$ & $-2.000000$ & $   9.85811819   -3.14159266 \mathrm{i} $ & $  -2.03366352+   3.07029505 \mathrm{i}$ \\ \hline
$A^{\rm tree}(1_{\bar{t}}^{+},2_{t}^{+},3_{\bar{q}}^{-},5_{g}^{-},4_{q}^{+})$ & & & $   0.02301119+   0.24677345\mathrm{i} $ \\
$r^{[1],a}(1_{\bar{t}}^{+},2_{t}^{+},3_{\bar{q}}^{-},5_{g}^{-},4_{q}^{+})$ & $-1.000000$ & $   4.97232920   $ & $   6.62000005   -2.34217273 \mathrm{i}$ \\ \hline
$A^{\rm tree}(1_{\bar{t}}^{+},2_{t}^{+},5_{g}^{-},4_{q}^{+},3_{\bar{q}}^{-})$ & & & $  -0.01891288   -0.23183261\mathrm{i} $ \\
$r^{[1],a}(1_{\bar{t}}^{+},2_{t}^{+},5_{g}^{-},4_{q}^{+},3_{\bar{q}}^{-})$ & $-2.000000$ & $   8.74792056   -3.14159267 \mathrm{i} $ & $   0.99823846+   2.11987447 \mathrm{i}$ \\ \hline
$A^{\rm tree}(1_{\bar{t}}^{+},2_{t}^{+},4_{q}^{+},3_{\bar{q}}^{-},5_{g}^{-})$ & & &  $   0.01041497   -0.09356833\mathrm{i}$ \\
$r^{[1],a}(1_{\bar{t}}^{+},2_{t}^{+},4_{q}^{+},3_{\bar{q}}^{-},5_{g}^{-})$ & $-2.000000$ & $   9.97517201   -3.14159267 \mathrm{i} $ & $  -2.81948074+   3.48650797 \mathrm{i}$ \\ \hline
$A^{\rm tree}(1_{\bar{t}}^{+},2_{t}^{+},4_{q}^{+},5_{g}^{-},3_{\bar{q}}^{-})$ & & & $   0.02301119+   0.24677345\mathrm{i} $ \\
$r^{[1],a}(1_{\bar{t}}^{+},2_{t}^{+},4_{q}^{+},5_{g}^{-},3_{\bar{q}}^{-})$ & $-1.000000$ & $   8.29930754 $ & $   0.22050369+   0.92161000 \mathrm{i}$ \\ \hline
$A^{\rm tree}(1_{\bar{t}}^{+},5_{g}^{-},2_{t}^{+},4_{q}^{+},3_{\bar{q}}^{-})$ & & & $  -0.01451328+   0.07862749\mathrm{i} $ \\
$r^{[1],a}(1_{\bar{t}}^{+},5_{g}^{-},2_{t}^{+},4_{q}^{+},3_{\bar{q}}^{-})$ & $-1.000000$ & $   8.29930751   $ & $  -4.93230098   -1.64974568 \mathrm{i}$\\ \hline \hline
$A^{\rm tree}(1_{\bar{t}}^{+},5_{g}^{-},4_{q}^{+},3_{\bar{q}}^{-},2_{t}^{+})$ & & & $   0.03342616+   0.15320512\mathrm{i}  $ \\
$r^{[1],b}(1_{\bar{t}}^{+},5_{g}^{-},4_{q}^{+},3_{\bar{q}}^{-},2_{t}^{+})$ & $ 0.000000$ & $   2.73111612   -3.22388847 \mathrm{i} $ & $   8.87318696+   3.05082436 \mathrm{i}$ \\ \hline
$A^{\rm tree}(1_{\bar{t}}^{+},4_{q}^{+},5_{g}^{-},3_{\bar{q}}^{-},2_{t}^{+})$ & & & $  -0.02301119   -0.24677345\mathrm{i} $ \\
$r^{[1],b}(1_{\bar{t}}^{+},4_{q}^{+},5_{g}^{-},3_{\bar{q}}^{-},2_{t}^{+})$ & $ 0.000000$ & $   2.73111596   -3.22388845 \mathrm{i} $ & $   8.48463218+   2.47700393 \mathrm{i}$ \\ \hline
$A^{\rm tree}(1_{\bar{t}}^{+},4_{q}^{+},3_{\bar{q}}^{-},5_{g}^{-},2_{t}^{+})$ & & &  $   0.00409831+   0.01494084\mathrm{i} $ \\
$r^{[1],b}(1_{\bar{t}}^{+},4_{q}^{+},3_{\bar{q}}^{-},5_{g}^{-},2_{t}^{+})$ & $ 0.000000$ & $   2.73111596   -3.22388844 \mathrm{i} $ & $   6.86621945   -1.31407912 \mathrm{i}$ \\ \hline
$A^{\rm tree}(1_{\bar{t}}^{+},4_{q}^{+},3_{\bar{q}}^{-},2_{t}^{+},5_{g}^{-})$ & & &  $  -0.01451328+   0.07862749\mathrm{i} $ \\
$r^{[1],b}(1_{\bar{t}}^{+},4_{q}^{+},3_{\bar{q}}^{-},2_{t}^{+},5_{g}^{-})$ & $-1.000000$ & $   4.47201710   -6.28323623 \mathrm{i} $ & $  10.06125557+   8.03168187 \mathrm{i}$ \\ \hline \hline
$A^{\rm tree}(1_{\bar{t}}^{+},2_{t}^{+},3_{\bar{q}}^{-},5_{g}^{-},4_{q}^{+})$ & & & $   0.02301119+   0.24677345\mathrm{i}  $ \\
$r^{[1],c}(1_{\bar{t}}^{+},2_{t}^{+},3_{\bar{q}}^{-},5_{g}^{-},4_{q}^{+})$ & $-2.000000$ & $   2.28510399   $ & $   2.72655473+   4.84327871 \mathrm{i}$ \\ \hline
$A^{\rm tree}(1_{\bar{t}}^{+},2_{t}^{+},5_{g}^{-},3_{\bar{q}}^{-},4_{q}^{+})$ & & & $  -0.00409831   -0.01494084\mathrm{i}  $ \\
$r^{[1],c}(1_{\bar{t}}^{+},2_{t}^{+},5_{g}^{-},3_{\bar{q}}^{-},4_{q}^{+})$ & $-1.000000$ & $   1.99739996   -3.14159265 \mathrm{i} $ & $   4.68165727+   2.92184337 \mathrm{i}$ \\ \hline
$A^{\rm tree}(1_{\bar{t}}^{+},5_{g}^{-},2_{t}^{+},3_{\bar{q}}^{-},4_{q}^{+})$ & & & $   0.01451328   -0.07862749\mathrm{i}  $ \\
$r^{[1],c}(1_{\bar{t}}^{+},5_{g}^{-},2_{t}^{+},3_{\bar{q}}^{-},4_{q}^{+})$ & $-1.000000$ & $   1.99739996   -3.14159266 \mathrm{i} $ & $   0.56499892+   6.27501703 \mathrm{i}$ \\ \hline
$A^{\rm tree}(1_{\bar{t}}^{+},2_{t}^{+},3_{\bar{q}}^{-},4_{q}^{+},5_{g}^{-})$ & & & $  -0.03342616   -0.15320512\mathrm{i} $ \\
$r^{[1],c}(1_{\bar{t}}^{+},2_{t}^{+},3_{\bar{q}}^{-},4_{q}^{+},5_{g}^{-})$ & $-1.000000$ & $   1.99739996   -3.14159265 \mathrm{i} $ & $   2.76146380+   3.49809487 \mathrm{i}$ \\ \hline
\end{tabular}
\end{center}
\caption{Examples of $A_L^{[1]}$ amplitudes contributing to the process $q_3 +\bar q_4  \to t \bar t + g_5 $.
The tree level amplitudes $A^{\rm tree}$ are given in units of $(100\, \mathrm{GeV})^{-1}$.}
\label{table:nterm3}
\end{table}

\newpage

\begin{table}[h]
\begin{center}
\begin{tabular}{| c|c|c|c| }
\hline
Helicity amplitude&$\;1/\epsilon^2\;$&$1/\epsilon$&$\epsilon^0$\\ \hline
$A^{\rm tree}(1_{\bar{t}}^{+},2_{t}^{+},3_{\bar{q}}^{-},4_{q}^{+},5_{g}^{-})$ & & & $  -0.03342616   -0.15320512\mathrm{i} $ \\
$r^{[1/2,q]}(1_{\bar{t}}^{+},2_{t}^{+},3_{\bar{q}}^{-},4_{q}^{+},5_{g}^{-})$ & $ 0.000000$ & $  -0.66666667 $ & $   0.14254796   -2.21798474 \mathrm{i}$ \\ \hline
$A^{\rm tree}(1_{\bar{t}}^{+},5_{g}^{-},2_{t}^{+},3_{\bar{q}}^{-},4_{q}^{+})$ & & & $   0.01451328   -0.07862749\mathrm{i} $ \\
$r^{[1/2,q]}(1_{\bar{t}}^{+},5_{g}^{-},2_{t}^{+},3_{\bar{q}}^{-},4_{q}^{+})$ & $ 0.000000$ & $  -0.66666667  $ & $   1.22048886   -2.09439510 \mathrm{i}$ \\  \hline
$A^{\rm tree}(1_{\bar{t}}^{+},2_{t}^{+},5_{g}^{-},3_{\bar{q}}^{-},4_{q}^{+})$ & & &  $  -0.00409831   -0.01494084\mathrm{i} $ \\
$r^{[1/2,q]}(1_{\bar{t}}^{+},2_{t}^{+},5_{g}^{-},3_{\bar{q}}^{-},4_{q}^{+})$ & $ 0.000000$ & $  -0.66666667  $ & $   0.41410458   -2.32237971 \mathrm{i}$ \\ \hline
$A^{\rm tree}(1_{\bar{t}}^{+},2_{t}^{+},3_{\bar{q}}^{-},5_{g}^{-},4_{q}^{+})$ & & & $   0.02301119+   0.24677345\mathrm{i} $ \\
$r^{[1/2,q]}(1_{\bar{t}}^{+},2_{t}^{+},3_{\bar{q}}^{-},5_{g}^{-},4_{q}^{+})$ & $ 0.000000$ & $  -0.66666667 $ & $   0.48191832   -2.09439510 \mathrm{i}$ \\ \hline \hline
$A^{\rm tree}(1_{\bar{t}}^{+},2_{t}^{+},3_{\bar{q}}^{-},4_{q}^{+},5_{g}^{-})$ & & & $  -0.03342616   -0.15320512\mathrm{i} $ \\
$r^{[1/2,t]}(1_{\bar{t}}^{+},2_{t}^{+},3_{\bar{q}}^{-},4_{q}^{+},5_{g}^{-})$ & $ 0.000000$ & $  -0.66666667 $ & $  -0.48762813   -2.07514502 \mathrm{i}$ \\ \hline
$A^{\rm tree}(1_{\bar{t}}^{+},5_{g}^{-},2_{t}^{+},3_{\bar{q}}^{-},4_{q}^{+})$ & & & $   0.01451328   -0.07862749\mathrm{i} $ \\
$r^{[1/2,t]}(1_{\bar{t}}^{+},5_{g}^{-},2_{t}^{+},3_{\bar{q}}^{-},4_{q}^{+})$ & $ 0.000000$ & $  -0.66666667 $ & $   1.08428102   -2.08237684 \mathrm{i}$ \\ \hline
$A^{\rm tree}(1_{\bar{t}}^{+},2_{t}^{+},5_{g}^{-},3_{\bar{q}}^{-},4_{q}^{+})$ & & & $  -0.00409831   -0.01494084\mathrm{i} $ \\
$r^{[1/2,t]}(1_{\bar{t}}^{+},2_{t}^{+},5_{g}^{-},3_{\bar{q}}^{-},4_{q}^{+})$ & $ 0.000000$ & $  -0.66666667 $ & $  -0.26372154   -2.56986449 \mathrm{i}$ \\ \hline
$A^{\rm tree}(1_{\bar{t}}^{+},2_{t}^{+},3_{\bar{q}}^{-},5_{g}^{-},4_{q}^{+})$ & & & $   0.02301119+   0.24677345\mathrm{i} $\\
$r^{[1/2,t]}(1_{\bar{t}}^{+},2_{t}^{+},3_{\bar{q}}^{-},5_{g}^{-},4_{q}^{+})$ & $ 0.000000$ & $  -0.66666667$ & $   0.00939493   -1.97232658 \mathrm{i}$ \\ \hline
\end{tabular}
\end{center}
\begin{center}
\caption{Examples of $A_L^{[1/2]}$ amplitudes  contributing to the process $q_3 +\bar q_4  \to t \bar t + g_5 $.
The tree level amplitudes $A^{\rm tree}$ are given in units of $(100\, \mathrm{GeV})^{-1}$.}
\end{center}

\label{table:nterm3}
\end{table}

\end{document}